\documentclass{aastex}
\usepackage{emulateapj5}
\usepackage{epsfig}
\slugcomment{To Appear in the Astrophysical Journal}
\shorttitle{Age-Dating ULIRG Mergers}
\shortauthors{Murphy et al.}

\begin{document}

\title{Age-Dating Ultraluminous Infrared Galaxies Along the Merger Sequence}

\author{T. W. Murphy, Jr.\altaffilmark{1}, B. T. Soifer\altaffilmark{2},
K. Matthews}

\affil{Palomar Observatory, California Institute of Technology, 320-47,
Pasadena, CA 91125}

\email{tmurphy@phys.washington.edu, bts@mop.caltech.edu, kym@caltech.edu}

\and

\author{L. Armus}

\affil{SIRTF Science Center, California Institute of Technology, 314-6,
Pasadena, CA 91125}

\altaffiltext{1}{Now at the University of Washington, Dept. of Physics,
Box 351560, Seattle, WA 98195}

\altaffiltext{2}{Also at the SIRTF Science Center, California Institute
of Technology, 314-6, Pasadena, CA 91125}

\begin{abstract}
Imaging spectroscopy using the new Palomar Integral Field Spectrograph
is presented for the Pa\( \alpha  \) line in four ultraluminous infrared
galaxies. The resulting integral field datacubes reveal line emission
possessing a wide range of complex spatial morphologies, often quite
different from the appearance of the continuum.  The velocity fields
are equally diverse in nature, often failing to resemble typical modes
of galactic motion.

We see a variety of interesting phenomena in the individual mergers
including star formation rates of 2--5 \( M_{\odot } \)~yr\( ^{-1} \)
in young tidal tails; a post-encounter disk which obeys the Tully-Fisher
relation; a large scale emission line nebula possibly associated with
a massive outflow; an apparently single merging system possessing two
distinct kinematical axes belying the presence of a second galaxy,
mostly obscured by its merging companion; and possible formation of
tidal dwarf galaxies.

In most cases, we are able to establish the geometry of the merger, and
thus estimate the time in the merger process at which we are viewing
the system.  The resulting range in estimated ages, some of which are
very young encounters (\( \sim 5\times 10^{7} \) yr), is not predicted
by merger models, which produce high rates of star formation either
1--\( 2\times 10^{8} \) years after the first encounter or very late
(\( \sim 10^{9} \) yr) in the merger process.  Even in the very young
mergers, despite a sample selection based on extended line emission, the
ultraluminous activity appears to be centrally concentrated on the nucleus
of one of the progenitor galaxies---namely the galaxy with a prograde
orbital geometry. The inferred extinction to these concentrations is high,
usually at least 1 magnitude at the wavelength of Pa\( \alpha  \).

The presence of a significant population of very young ultraluminous
mergers, together with the majority of ultraluminous infrared galaxies
existing in the final stages of merger activity, indicates that the
ultraluminous galaxy phase is at least bimodal in time. An evolutionary
scenario is proposed for ultraluminous infrared galaxies, wherein the
far-infrared luminosity may undergo multiple ultraluminous bursts during
the course of the encounter. A substantial fraction of the merger
lifetime may be spent in a phase identified with the less powerful
luminous infrared galaxy class.
\end{abstract}


\newcommand{\rubble}{IRAS~01521$+$5224}

\newcommand{\double}{IRAS~10190$+$1322}

\newcommand{\bubble}{IRAS~17574$+$0629}

\newcommand{\trouble}{IRAS~20046$-$0623}

\newcommand{\pa}{Pa$\alpha$}

\section{Introduction}

Ultraluminous infrared galaxies (ULIRGs) are among the most luminous
sources in the universe, with infrared luminosities of \( L_{ir}\ga
10^{12}L_{\odot } \).  Imaging surveys of ULIRGs find that the vast
majority of these systems are morphologically distorted, with spatial
structures indicative of galactic mergers \citep{sand88,twm96,clements}.
ULIRGs are found to exist in various states of merging, from well
separated galaxies to single nucleus systems that appear to have completed
their nuclear coalescence. From millimeter wavelength studies of molecular
gas in ULIRGs, it is found that the constituent galaxies are typically
large, gas-rich spirals \citep{sandsco,downes,solomon}. In general
terms, the merging process destabilizes the orbits of gas within the
galaxies, leading to accumulation of gas in the central potential of the
individual galaxies or merger remnant. The resulting high concentrations
of molecular gas stimulate the production of stars, or in some cases fuel
a massive central black hole, though both can also occur simultaneously.
In addition, cloud-cloud collisions may occur within the disrupted
disk, or as the galaxies begin to overlap, leading to widespread star
formation. Veiled by the vast quantities of dust mixed with the molecular
gas, these energetic processes are seen via dust-reprocessed light at
far-infrared wavelengths, and are held to be responsible for the extreme
luminosity of these systems \citep[see][for a review]{review}.

A new integral field spectrograph \citep{pifs} working at near-infrared
wavelengths on the Palomar 200-inch Telescope has provided a new
way to probe the properties of the interstellar medium in these
galaxies. Combining imaging capabilities with longslit spectroscopy,
the integral field data permit simultaneous acquisition of spatial and
velocity information of emission line gas over a two-dimensional region
on the sky. For complex sources such as ULIRGs, the advantage over more
traditional longslit or narrow-band techniques is substantial.

This paper presents the spatial and kinematic structures of the
Pa\( \alpha  \) line emission nebulae in four ULIRGs of various
morphological types. Most of the line emission observed in these
galaxies is spectrally unresolved at the current resolution. Therefore,
this paper concentrates on the morphologies and velocity fields
of the line emission structures. The four galaxies are taken
from the 2 Jy sample of \citet{str90,str92}, and are selected for
their peculiar spatial and velocity structures in line emission
as seen in two-dimensional near-infrared spectra taken from
\citet{twm99,twm01}. Table~\ref{tab:ulirgs} lists the properties of the
four ULIRGs in the present sample. All cosmology-dependent calculations
in this paper assume \( H_{0}=75 \) km~s\( ^{-1} \)~Mpc\( ^{-1} \) and \(
q_{0}=0 \).

\section{Observations and Data Reduction}

\subsection{Integral Field Data\label{pifsreduc}}

Observations were made using the Palomar Integral Field Spectrograph
(PIFS) situated at the \( f \)/70 Cassegrain focus of the 200-inch
Hale Telescope.  A description of this instrument along with general
observing procedures can be found in \citet{pifs}. In brief,
the integral field spectrograph delivers simultaneous spectral
information across a contiguous two-dimensional field of view with
essentially seeing-limited spatial resolution in both dimensions.
For these observations, the 5\farcs 4\( \times  \)9\farcs 6 field
of view was rotated to optimally cover the extent of the galaxies as
seen in broadband infrared light, with two contiguous field positions
required to cover the larger structure of \rubble . All observations
were made in clear conditions. Table~\ref{tab:pifs} provides a summary
of the observations. The \( R\approx 1300 \) resolution mode was
used to obtain spectra centered on the redshifted Pa\( \alpha  \)
line in each galaxy. Separate sky exposures were alternated with
the on-source integrations, with individual integration times of
300~s. A positional dither pattern was employed for the sequence of
integrations enabling recovery of seeing-limited spatial resolution
in the cross-slit direction. Observations of nearby stars for the
purpose of evaluating the point spread function (PSF) accompanied
the spectral observations. Wavelength calibration is provided through
a combination of OH airglow lines and arc lamp spectra taken at the
time of observation. Atmospheric opacity and spectral flat-fielding are
compensated simultaneously using the light from a G dwarf star, spread
uniformly across the field of view.

Data reduction consists of subtracting the sky integrations,
interpolating static bad pixels and cosmic ray artifacts, dividing
by the G star spectrum, and multiplying by a blackbody matched to
the G star's temperature. Spatial and spectral distortions are then
corrected using previously generated distortion maps appropriate for
the particular grating setting. Co-registration of the slits in the
spatial dimension is based upon observation of the G star with its light
extended perpendicular to the slit pattern by chopping the telescope
secondary mirror in a triangle-wave pattern. The two-dimensional spectra
from the eight slits are placed into a three-dimensional datacube
according to the positional dither pattern, with a common wavelength
axis established by the calibration lines. Residual OH airglow lines
are removed by subtracting a scaled version of the raw sky spectrum,
with typical scalings of \( \sim  \)2\% in absolute value, sometimes as
large as 10\%. Photometric variability among individual integrations is
compensated by small scaling adjustments such that the object flux is
consistent from one integration to the next. A more detailed description
of these general procedures may be found in \citet{pifs}.

In the data presented here, the pure continuum images are constructed
directly from the integral field datacube, avoiding spectral regions
associated with OH airglow emission or poor atmospheric transmission. For
each spatial pixel, a linear fit is made to the line-free portion of
the continuum spectrum. The line images are formed by subtracting this
continuum fit from the datacube, then summing in the spectral dimension
over a range encompassing the line emission.  Because they are extracted
from the same datacube, the co-registration of the continuum and line
images is implicit.

The velocity fields are constructed in the following manner. At each
spatial pixel, the wavelength of the peak emission is computed, and
converted to a velocity.  Because spatial gradients of flux across the
individual slits can mimic shifts in velocity, the morphology of the
line emission at the line peak is used to assess these gradients and
correct the peak velocity values. These corrections are typically no
larger than 30--40 km~s\( ^{-1} \). A ``maximum Pa\( \alpha  \)'' image
is also presented for each ULIRG in order to show weaker emission with
greater sensitivity. This image, also continuum subtracted, represents
a single spectral resolution element centered for each spatial pixel on
the wavelength of maximum line emission intensity. In other words, this
image extracts spectral information only around the peak line intensity,
thereby following the velocity indicated in the velocity field.

Photometric calibration was performed using the photometric system
of \citet{persson}.  Images of the standard stars were taken in the
spectrograph's imaging mode through a \( K_{s} \) filter, and the total
flux was compared to images of the science object taken in the same
manner. The photometry is corrected for atmospheric opacity, and for
Galactic extinction based on dust measurements by \citet{schlegel},
and using the extinction law of \citet{extinction}. For each of the
galaxies in this sample, near-infrared spectra \citep[e.g.,][]{twm01}
were available from which to judge the continuum shape, so that the
continuum flux density at Pa\( \alpha  \) could be related to the
continuum flux density in the center of the \( K_{s} \) bandpass. The
photometric calibration was supplemented by similar procedures using
the \( K_{s} \) imaging discussed below.

\subsection{Supplementary Imaging}

Continuum and narrow-band H\( \alpha + \){[}\ion{N}{2}{]} images accompany
the integral field data in Section~\ref{results}, where available, in
order to provide a better sense of the overall structure of the continuum
and line emission light. Most of the near-infrared images were taken
through the wide-open 10 arcsecond slit of the Palomar Longslit Infrared
Spectrograph \citep{larkin}, operating on the 200-inch Telescope, and
using a \( 256\times 256 \) HgCdTe array with a pixel scale of 0\farcs
167 pixel\( ^{-1} \). The near-infrared image of \trouble\ was obtained
using a \( 58\times 62 \) InSb array camera also operating on the
200-inch Telescope with a pixel scale of 0\farcs 313 pixel\( ^{-1} \).
Table~\ref{tab:hna} presents the parameters of these observations,
all of which used \( K_{s} \) filters.

Visual imaging is also presented for the four galaxies in this paper,
with \( r \) band and H\( \alpha + \){[}\ion{N}{2}{]} images obtained
at the Palomar 60-inch Telescope. For \bubble\ and \trouble , \( r \)
band images are presented in \citet{twm96}, and duplicated here. The
data reduction procedures for the \( r \) band images in this paper
are described in \citeauthor{twm96} The H\( \alpha + \){[}\ion{N}{2}{]}
images were taken with narrow band filters \( \sim  \)10 nm wide, using
the \( r \) band images as the continuum reference. Data reduction for
the narrow band images is similar to that for the \( r \) band images,
followed by scaling, shifting, rotating, and subtracting the \( r \) band
continuum images such that nearby stars were nulled on average. The \(
r \) filter bandpass does contain the H\( \alpha  \) light, but generally
at < 10\% of the total flux.  Photometric calibration was accomplished
by use of the spectrophotometric standards from \citet{oke}, correcting
for atmospheric opacity, and for Galactic extinction by the same methods
described at the end of Section~\ref{pifsreduc}. Table~\ref{tab:p60}
presents a summary of the visual observations. No H\( \alpha  \) data
exist for \double , and the \( r \) band image for this galaxy was taken
through clouds with the telescope unguided.

\section{Individual Objects\label{results}}

\subsection{IRAS~01521\protect\( +\protect \)5224\label{iras0152}}

\subsubsection{Morphology of Continuum \& Line Emission\label{morph0152}}

\begin{figure*}[tbh]
\centerline{\epsfig{file=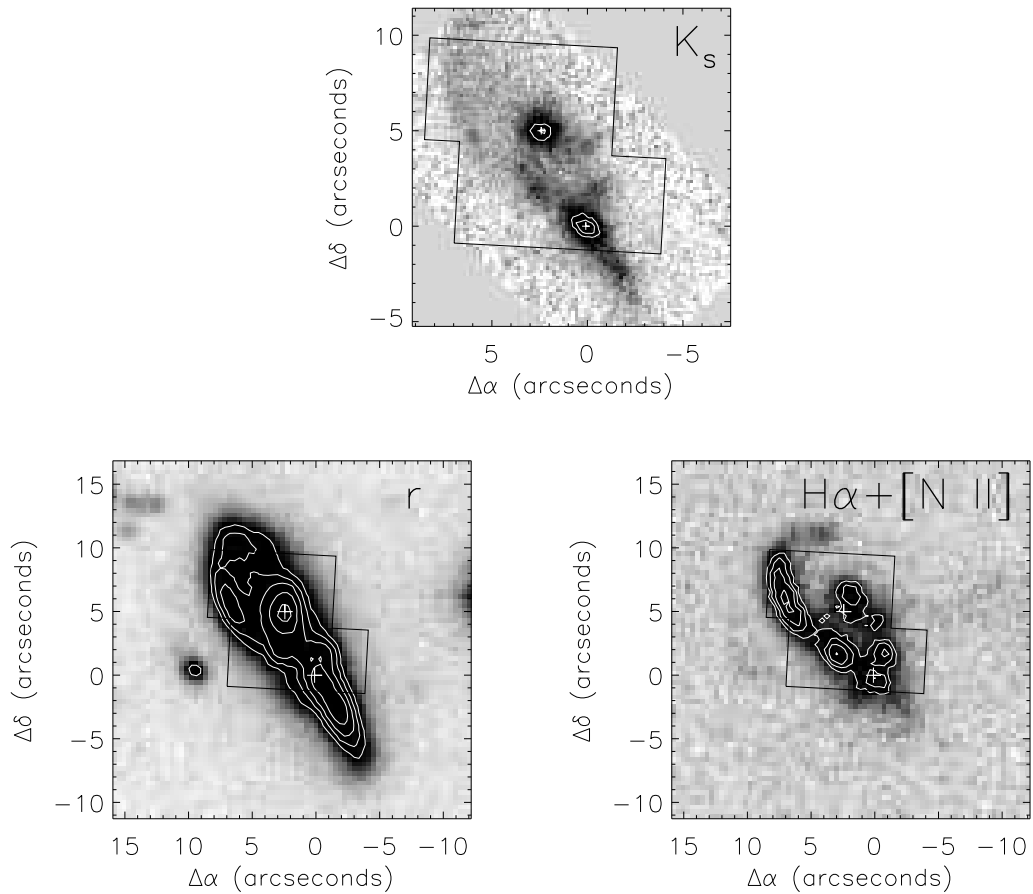,width=6.0in}}
\caption{\label{fig:0152cont}Continuum images of IRAS~01521$+$5224 
in $K_{s}$ (top) and $r$ (lower-left) bands. These images
portray a dramatically disturbed ULIRG, with four $r$ band peaks
visible, two of which are the obvious nuclei in the infrared image. Note
the lack of any major features extending beyond the main grouping. A
narrow band image in H$\alpha +${[}\ion{N}{2}{]} is presented at
right, showing many emission peaks arranged along a ring-like structure.
The irregular box outline shows the region observed with the integral
field spectrograph. Crosses indicate the positions of the near-infrared
continuum peaks. North is up, and east to the left.}
\end{figure*}

\rubble\ is a morphologically complex galaxy consisting of two
primary galaxies and many clumps of distributed tidal debris. The
two nuclei are separated by 5\farcs 5 (7.6 kpc projected separation)
at a position angle of 25\( ^{\circ } \), as determined from the
near-infrared image. Near-infrared spectra of this galaxy appear typical
of starburst-dominated ULIRGs \citep{twm01}. Figure~\ref{fig:0152cont}
shows both a \( K_{s} \) band image and an \( r \) band image of the
galaxy, along with an image in H\( \alpha + \){[}\ion{N}{2}{]}. The
nature of the lower surface brightness features is not immediately
clear from these images. These features could be tidal tails, small
companion galaxies, or fragments of the primary galaxies. The \( r \)
band image demonstrates that the high surface brightness features are
contained within a relatively small region. In other words, there is no
obvious tidal tail component extending away from the main concentration
of light at the current level of image depth. Comparing to simulations of
galaxy mergers \citep[e.g.,][]{hos96, barnes96}, one finds a similarity
between this morphologically complex system and those model systems
as seen shortly after the first close encounter. Note the absence of a
distinct southern nucleus in the \( r \) band image, in contrast to the
appearance of a well defined nucleus in the \( K_{s} \) image. The H\(
\alpha + \){[}\ion{N}{2}{]} image, which has a dramatically different
morphology from the continuum emission, is discussed in relation to the
Pa\( \alpha  \) images in Section~\ref{sfr0152}.

\begin{figure*}[tbh]
\centerline{\epsfig{file=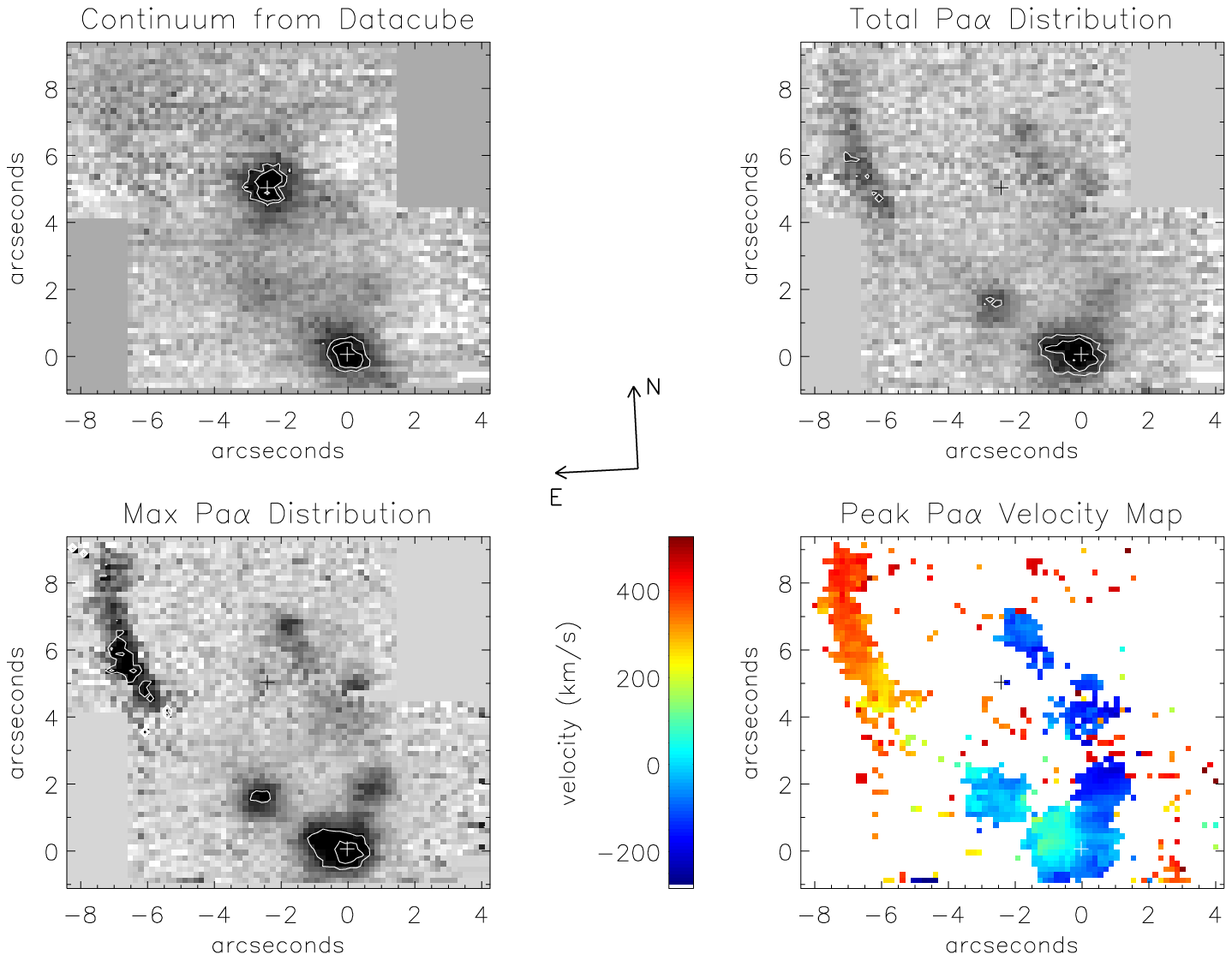,width=6.0in}}
\caption{\label{fig:0152pifs}PIFS data on the central region
of IRAS~01521$+$5224.  At top left is the line-free continuum
image reconstructed from the datacube.  At top right is the total
Pa$\alpha $ emission, continuum subtracted, and summed from $-325$
to $+480$ km~s$^{-1}$ relative to the systemic velocity. A
higher contrast Pa$\alpha$ image, tracing the peak Pa$\alpha$ 
emission as described in the text, is shown at lower left. Finally,
the velocity field is represented at lower right as a color diagram,
with red representing redshifted gas, and blue representing blueshifted
gas. The velocity scale is indicated to the left of the plot. In all
images, the continuum peak locations are indicated by crosses. Field
orientation is indicated by the arrows, with ``left'' corresponding to
a position angle of 87$^{\circ }$.}
\end{figure*}

Figure~\ref{fig:0152pifs} presents the PIFS data of \rubble , showing
images of the pure continuum, total Pa\( \alpha  \) line emission,
maximum Pa\( \alpha  \) emission, and the velocity field of the Pa\(
\alpha  \) emission peak. The large extent of this galaxy required
observations in two adjacent and contiguous fields, the positions of
which are indicated in Figure~\ref{fig:0152cont}.  The Pa\( \alpha  \)
emission distribution is markedly different from that of the continuum
light. The Pa\( \alpha  \) peak is coincident with the southern galaxy
continuum peak, though it is not symmetrically distributed about the
continuum peak position. The northern galaxy nucleus has very little
Pa\( \alpha  \) emission associated with it, identifying the southern
nucleus as the primary source of star formation, and thus presumably of
the far-infrared emission.

The off-nuclear Pa\( \alpha  \) emission is extended over an impressive
15 kpc, mostly arranged in clumps along a ring-like structure---much
like that seen in the H\( \alpha  \) image. Combined with the velocity
information provided in the integral field datacube, the morphology of
these features enables one to develop a model for the current merger
geometry of the system, as will be discussed in Section~\ref{merger0152}.

\subsubsection{Extinction \& Star Formation\label{sfr0152}}

Using the H\( \alpha  \) image together with the Pa\( \alpha  \) image,
one may compute line ratios to derive measures of the extinction to the
various regions in the galaxy. The total Pa\( \alpha  \) flux within
the PIFS field is \( \gtrsim 2.1\times 10^{-17} \) W~m\( ^{-2} \),
compared to \( 1.3\times 10^{-16} \) W~m\( ^{-2} \) for H\( \alpha +
\){[}\ion{N}{2}{]} in a similar aperture.  Assuming an intrinsic H\(
\alpha / \)Pa\( \alpha  \) line ratio of 8.6 \citep[][Case B, $n_e=10^4$
cm$^{-3}$, $T=10000$ K]{ost}, a lower limit can be placed on the average
extinction to the source of \( A_{V}>0.5 \) mag. The extinction at Pa\(
\alpha  \) is interpolated from the extinction law of \citet{extinction},
with \( A_{Pa\alpha }=0.145A_{V} \). This and all other extinction
measures in this paper assume no contribution from {[}\ion{N}{2}{]}, which
is often half the strength of H\( \alpha  \) in star forming regions.
Including a flux contribution from {[}\ion{N}{2}{]} at this level adds
another half-magnitude to the estimated visual extinction. As such,
all extinction values are stated as lower limits. The southern nucleus
itself contributes 30\% of the total Pa\( \alpha  \) flux, though a quick
inspection of Figure~\ref{fig:0152cont} shows that the H\( \alpha  \) flux
from this nucleus contributes a much smaller fraction of the total H\(
\alpha  \) line emission. The extinction to the southern nucleus line
emission measures \( A_{V}\gtrsim 2.5 \) mag. Away from the southern
nucleus, the visual extinction is typically 0--0.75 mag.

The hydrogen recombination flux can be used to calculate the star
formation rate (SFR) in the galaxy, following the prescription
for converting H\( \alpha  \) flux to the global SFR outlined in
\citet{kennicutt}. The Pa\( \alpha  \) line is used rather than
the H\( \alpha  \) line, as it suffers much less from extinction.
An extinction-corrected H\( \alpha  \) flux, used in calculating the
SFR, is inferred from the Pa\( \alpha  \) flux using an intrinsic line
ratio of 8.6, as introduced above. The Pa\( \alpha  \) flux is itself
uncorrected for extinction, such that the SFR estimate performed in this
manner represents a lower limit.  For the total Pa\( \alpha  \) emission
recorded in the PIFS field, the SFR is computed to be \( \gtrsim 20 \)
\( M_{\odot } \)~yr\( ^{-1} \). As suggested above, the southern nucleus
is responsible for about 30\% of this total, with the rest split between
the isolated clumps seen in Figure~\ref{fig:0152pifs}.

A surprisingly large amount of star formation appears to be occurring
in isolated clumps distributed along what are thought to be tidal
tails---discussed at length below. Tidal tails generally do contain gas,
as they are composed of material rather indiscriminately pulled from a
contiguous region of the parent galaxy, within which stars and gas are
mixed with similar global distributions. Star formation is known to occur
in tidal tails \citep[e.g.,][]{chromey}, and tails are, in fact, seen to
contain many young, blue stars \citep{tyson,russian}.  The star formation
rate estimated for the bright Pa\( \alpha  \) clump in the northeast
tail is \( \sim 6 \) \( M_{\odot } \)~yr\( ^{-1} \), which is unusually
high for such environments \citep{hibbard96,duc98,duc00}. Typical star
formation rates in tidal clumps and tidal dwarf galaxies are estimated at
0.05--0.2 \( M_{\odot } \) yr\( ^{-1} \). The highest previously observed
star formation rates in tidal tails also occurs in ULIRGs, with about 0.5
\( M_{\odot } \)~yr\( ^{-1} \) \citep{hos98}. The very high rate of star
formation in the clump may be the result of gas compression induced by
crossing orbits in the forming tail. The work by \citet{toomre} points
out that shortly after tidal tails develop, the tidal material on the
inside of the tail overtakes the outer material. The crossing orbits
that result compress the gas within the tail, leading to heightened star
formation---at least in low inclination systems \citep{wallin}. This
scenario, which is relevant only for very young tails, is consistent
with the estimated age and geometry of the \rubble\ encounter, as
discussed below.

\subsubsection{Merger Geometry\label{merger0152}}

The complex morphology of the \rubble\ merger does not lead to an
immediate understanding of the merger state of this galaxy pair. At
first glance, the H\( \alpha  \) and Pa\( \alpha  \) emission appears
to be distributed around an elliptical ring, roughly centered on the
northern galaxy nucleus. Though rings of comparable size and star
formation activity have been observed in galaxies like the Cartwheel
\citep{higdon1} and AM~0644\( - \)741 \citep{higdon2}, we favor an
interpretation that naturally accounts for the various observed properties
by way of tidal tails. We briefly discuss the ring interpretation before
presenting the tidal tail model. A more complete discussion of the
observations supporting the merger model presented below can be found
in \citet{thesis}.

\begin{figure*}[tbh]
\centerline{\epsfig{file=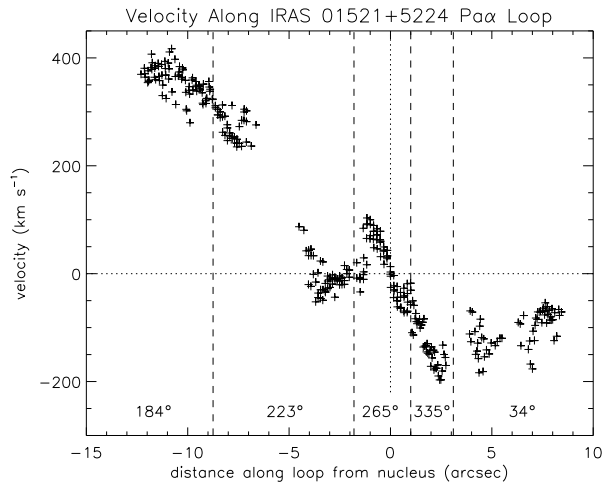,width=3.5in}}
\caption{\label{fig:0152rot}Segmented position-velocity plot
along the loop of Pa$\alpha$ emission in IRAS~01521$+$5224,
starting in the northeast, and winding around in a counter-clockwise
sense, through the southern continuum peak, then back to the north. The
loop was broken into five straight-line segments, with breakpoints
and position angles represented on the plot, delineated by dashed
lines. The displacement is plotted as accumulated linear distance
along the segments. The southern nucleus is at the zero position on
the horizontal axis, displaying a rotation curve with an amplitude of
about 150 km~s$^{-1}$.  Note the fast moving clump at upper left
(northeast clump), with a velocity gradient projecting directly away
from the southern nucleus center. This clump was broken into two linear
slit segments owing to its spatially curved nature.}
\end{figure*}

Arguments for a ring include the approximate elliptical shape of
the line emission, and the location of the northern galaxy near the
center of the ellipse. If the material within the ring is orbiting
the central galaxy, the maximum projected radial velocities will
be observed at either end of the ellipse---much like what we see
in Figure~\ref{fig:0152pifs}. Figure~\ref{fig:0152rot} shows a
position-velocity plot constructed along a spatial loop following
the Pa\( \alpha  \) line emission. From this it is seen that there is
little or no change in velocity along the western part of the ellipse,
arguing against the ring interpretation.  Moreover, the velocity
mismatch of \( \sim 200 \) km~s\( ^{-1} \) across the minor axis of
the ellipse is unexplained by the rotating ring model, since this model
predicts identical velocities across the ring's projected minor axis.
This mismatch could be explained in part by a radial expansion of the
rotating ring \citep[e.g.,][]{higdon97}. However, the asymmetry in the
velocity profile around the ellipse---most notably the monotonic western
portion---remains problematic.  Both the ring and tidal tail models
for \rubble\ require special viewing angles leading to coincidental
projections, projecting the southern nucleus either onto the ring or
onto the tail from the northern galaxy.

\begin{figure*}[tbh]
\centerline{\epsfig{file=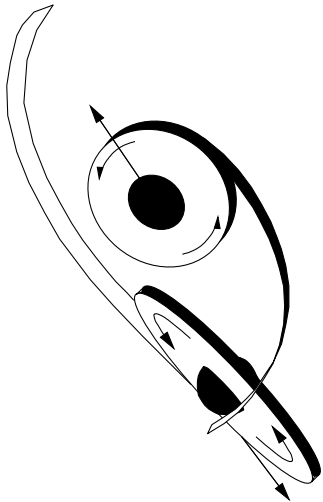,width=3.5in}}
\caption{\label{fig:0152model}A schematic model of the
IRAS~01521$+$5224 system, as seen from our viewpoint. The northern
(upper) galaxy, nearly face-on, is in the foreground of the southern
galaxy, which is seen nearly edge-on. The eastern tail spins off of
the southern galaxy, and is seen in almost pure recession.  The western
tail, extending from the northern galaxy, is almost in the plane of the
sky, explaining the very small observed velocity gradient along this
tail. Bulges are placed in the diagram to approximate the appearance of
the $K_{s}$ image in Figure~\ref{fig:0152cont}.}
\end{figure*}

\begin{figure*}[tbh]
\centerline{\epsfig{file=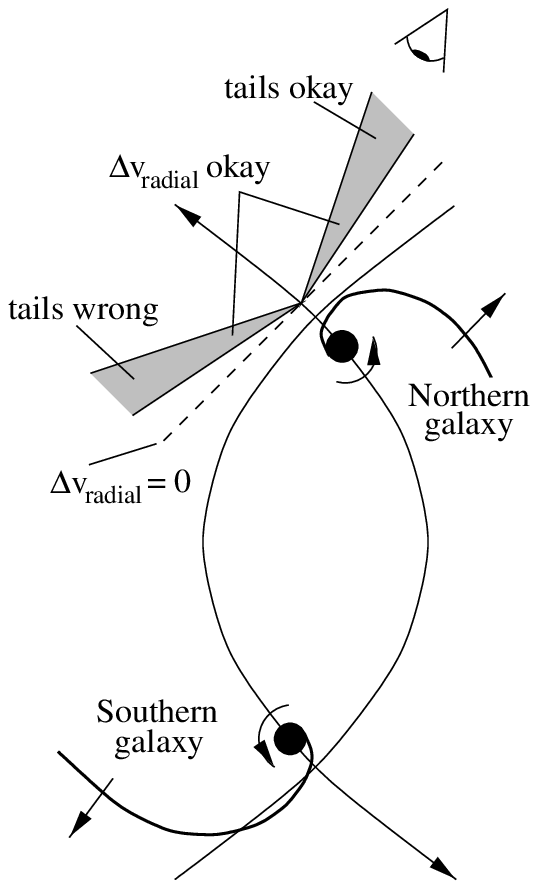,width=3.5in}}
\caption{\label{fig:0152geom}Schematic diagram of the likely
orbital plane geometry of the IRAS~01521$+$5224 merger, indicating probable
locations of our vantage point.  The dashed line represents a plane from
which the galaxies would currently show no relative radial velocity. The
shaded grey region, when revolved about the current relative velocity
vector, represents the region of space from which the northern galaxy
is slightly blueshifted relative to the southern galaxy.  Only from the
upper right portion of this region do the tidal tails have the correct
observed velocities.}
\end{figure*}

We favor the tidal tail model over the ring model, in part because
tails are ubiquitous features of merging systems, while rings are
far less common, requiring a carefully arranged encounter---often
with the disturbing galaxy passing through the central region of
the ringed galaxy from a polar direction. Figure~\ref{fig:0152model}
is a schematic representation of our view on \rubble\ in the context
of our model of the merger geometry. The orientation follows that in
Figure~\ref{fig:0152cont}.  In this model, the southern galaxy is in
the background, seen nearly edge-on.  Thus the elongated structure in
the \( K_{s} \) image at a position angle near 45\( ^{\circ } \) and
symmetrically located with respect to the southern nucleus represents the
disk of the southern galaxy. The edge-on appearance is also suggested
by the high nuclear extinction to the southern nucleus---both in
line emission and continuum light. The arcing feature to the east
and north---as seen in line emission---is a tidal tail connecting
to the back side of the southern galaxy disk plane. The northern
galaxy, viewed more nearly face-on, has a tenuous tail to the west
that is by chance projected onto the vicinity of the southern galaxy
nucleus. Figure~\ref{fig:0152geom} shows what this pair of galaxies might
look like from a vantage point directly over the orbital plane. The
geometry proposed for the southern galaxy closely resembles that seen
following the first encounter in many numerical merger models containing
prograde disks \citep[see especially][Fig.~13]{hos96}.

\begin{figure*}[tbh]
\centerline{\epsfig{file=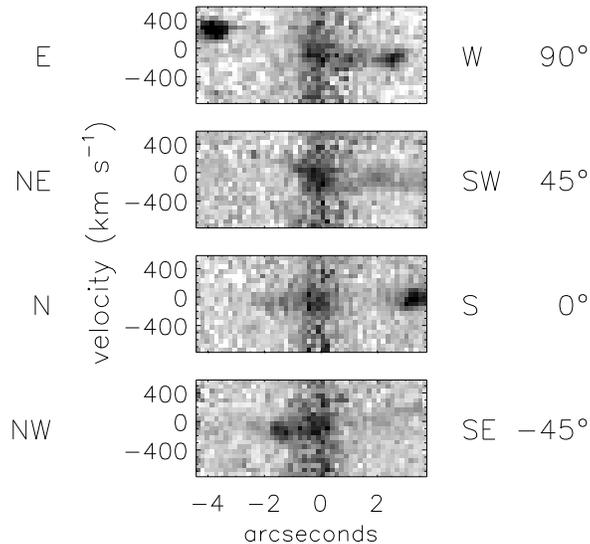,width=3.5in}}
\caption{\label{fig:0152nnuc}Four two-dimensional spectra of
the northern nucleus extracted from the IRAS~01521$+$5224 datacube
corresponding to four slits at position angles of 90$^{\circ }$,
45$^{\circ }$, 0$^{\circ }$, and $-45^{\circ }$ show that
there is Pa$\alpha$ emission associated with this nucleus. The
vertical scale is in km~s$^{-1}$ relative to the southern nucleus
velocity.  The continuum from the nucleus can be seen in the center of
each spectrum, with line emission appearing at about $-50$ km~s$^{-1}$.
A slight rotational motion is inferred from the apparent tilts
of the line emission, with the maximum gradient probably at a position
angle between 45$^{\circ }$ and 90$^{\circ }$.  The emission from
the western tail is seen to smoothly interface with the nuclear emission,
seen most clearly in the top panel.}
\end{figure*}

The senses of rotation depicted in Figures~\ref{fig:0152model}
and \ref{fig:0152geom} are derived from the Pa\( \alpha  \) velocity
field. Though the velocity field in Figure~\ref{fig:0152pifs} does not
yield information on the radial velocity or rotation state of the northern
nucleus, two-dimensional spectral extractions from the datacube, shown in
Figure~\ref{fig:0152nnuc}, provide enough information to determine that
the northern galaxy is blueshifted relative to the southern galaxy by \(
\sim  \)50 km~s\( ^{-1} \), and that it is weakly rotating at a position
angle between 45\( ^{\circ } \) and 90\( ^{\circ } \). An arrangement
such as that shown in Figure~\ref{fig:0152geom} is consistent with
the morphologies and velocity fields observed in \rubble , with our
view restricted to the shaded region, and very likely from a vantage
close to the orbital plane. The northern galaxy in this figure is not
accurately represented, and should be rotated out of the plane of the
paper in order to more closely match the observations.

The proposed geometry, along with the presumed vantage point from near the
orbital plane, means that the southern galaxy's rotation plane is nearly
coincident with the orbital plane, in a highly prograde configuration. The
northern galaxy is then inclined with respect to the orbital plane,
itself being weakly prograde \citep[see][for definitions of prograde
and retrograde]{toomre}.

The following observations lend support to the tidal tail
interpretation. The eastern tail very naturally interfaces with the
southern galaxy disk plane, both in terms of position angle on the sky
and the velocity field, which matches the rotation sense of the southern
galaxy. The western tail also joins the northern nucleus, seamlessly
tying into its velocity field, as indicated in Figure~\ref{fig:0152nnuc}.
The nearly constant velocity of the western tail is consistent with the
deduction that we view the northern galaxy nearly face-on, such that
the tail---moving primarily in the rotation plane of its parent galaxy
\citep{toomre}---has little motion along our line-of-sight. Though the
H\( \alpha  \) morphology is rather closely elliptical in shape, there
is a slight departure at the northern end of the eastern tail. While
it is true that the western tail coincidentally terminates on both the
northern and southern galaxies, the H\( \alpha  \) image hints that
the tail may slightly overshoot the southern nucleus, as reflected in
Figure~\ref{fig:0152model}.

Having adopted the tidal tail interpretation of the \rubble\ morphology,
a few other circumstances fall neatly into place. The geometry of
the model, as portrayed in Figure~\ref{fig:0152geom}, shows that the
two galaxies have just passed a close encounter. This explains the
impressive level of complexity seen in \rubble , and accounts for the
short, high-surface-brightness appearance of the newly formed tails. In
the same vein, the very high rate of star formation in the tails can be
understood in terms of gas compression resulting from crossing orbits in
young tails \citep{toomre,wallin}. This compression is relevant only for
low inclination systems, as we believe the southern nucleus in \rubble\ to
be. The highly prograde geometry of the southern galaxy not only explains
the prompt formation of a tidal tail, but also may explain \emph{why}
this galaxy is ultraluminous. Prograde galaxies are highly susceptible
to the formation of bar potentials around the nucleus, so that they are
very effective at funneling gas into high concentrations in the central
regions of the galaxy. It is this fast-acting response that enables
\rubble\ to appear ultraluminous at this early stage of merging. This
topic is discussed at greater length in Section~\ref{discussion}.

\subsubsection{Age of Merger\label{dwarf0152}}

The current position and velocity of the tidal tails can be used
to estimate a merger age for \rubble . The elongated Pa\( \alpha  \)
feature in the northeastern corner of the field is appreciably redshifted
relative to the southern nucleus.  This clump (hereafter the northeast
clump) appears to lie in the tidal tail originating from the southern
galaxy. The clump's velocity gradient, seen in Figure~\ref{fig:0152rot},
points directly at the plot origin, corresponding to the position-velocity
of the southern nucleus. Such an arrangement in position-velocity space
would be expected in a very simplified model of tidal tail production,
wherein a mass of material is extracted from a single location in the
source at an instant in time. In such a scenario, the faster moving
mass moves farther away in the same amount of time. In reality, tidal
tails are not formed instantly, and the constituent mass is pulled out
of a zone in the galaxy, rather than a single point. Yet overall, the
simulations of tidal tail formation by \citet{toomre} are in agreement
with this generic velocity profile.

The physical distance between the southern nucleus and the
middle of the northeast tidal clump is around 15 kpc, based on
the geometrical projections surmised from the geometry developed in
Section~\ref{merger0152}. This distance is traveled in \( 5\times 10^{7}
\) yr at the measured recessional speed of 300 km~s\( ^{-1} \), which is
roughly consistent with the estimated timescale for the formation of such
structures in the galactic merger models \citep[e.g.,][see in particular
panel 3 of Fig. 13]{barnes96,hos96}.  It was mentioned previously that
the high rate of star formation in young tidal tails is consistent with
the formation of caustics (i.e., crossing orbits) as found in simulations
of tidal disturbances. \citet{wallin} showed that the caustic within
a tidal tail starts at the root of the tail and travels outward along
the tail. This provides a natural explanation for the detachment of the
intense star formation region from the main body of the southern galaxy,
and can also be used to compare the development stage of the \rubble\
merger with that of merger simulations.

One may speculate that the northeastern tidal clump, with its high
velocity and high rate of star formation, may evolve into a detached,
blue galaxy with properties typical of dwarf galaxies. If the clump is
not tidally stable, this single feature could give rise to multiple dwarf
galaxies. The notion that dwarf galaxies can be spawned in tidal tails
was originally proposed by \citet{zwicky}.  \citet{schweizer} found
observational evidence for this conjecture at the tip of the southern
tidal tail in the Antennae (NGC~4038/9). More recent observations
\citep{mirabel,hibbard96,duc98} have found additional evidence for the
tidal clump/dwarf galaxy connection. Simulations verify the plausibility
of self-gravitating conglomerates with masses similar to those of dwarf
galaxies forming out of tidal tail material \citep{barnes92,elm}. The
clump is almost certainly tidally unstable, given its proximity to
the parent galaxies and very large velocity gradient. It is not clear
that even the outermost, fastest-moving material has enough velocity to
escape the system. Even if ultimately bound, this knot of material could
develop into a number of self-gravitating systems in very long-lived
orbits about the center of mass of the post-merger galaxy system. More
detail on the dynamical properties of the northeastern tidal clump can
be found in \citet{thesis}.

\subsection{IRAS~10190\protect\( +\protect \)1322}

\subsubsection{Morphology of Continuum \& Line Emission\label{morph1019}}

\begin{figure*}[tbh]
\centerline{\epsfig{file=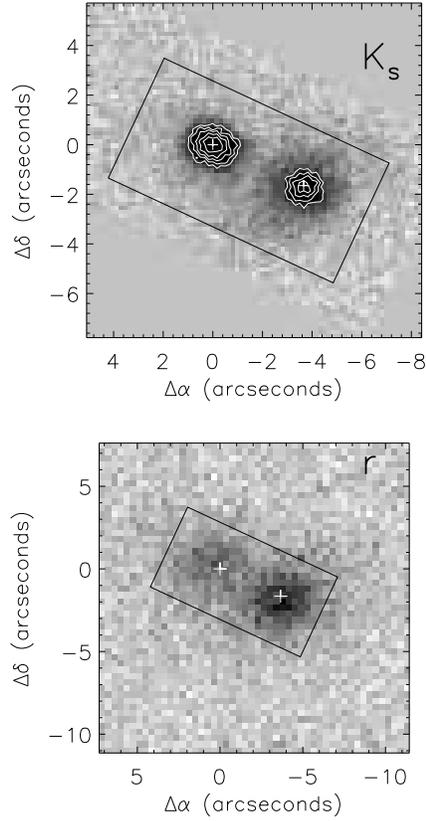,width=2.5in}}
\caption{\label{fig:1019cont}Continuum images of IRAS~1019$+$1322
in the $K_{s}$ and $r$ bands. The $r$ band image was
obtained in adverse conditions, preventing a meaningful investigation
of large scale tidal debris. Note the reversal of the dominant nucleus
between the visible and infrared images. The PIFS field is represented as
a rectangular outline in both images. Crosses indicate the positions of
the near-infrared continuum peaks. North is up, and east is to the left.}
\end{figure*}

\double\ is a double-nucleus ULIRG with a projected nuclear separation
of 4\farcs 0 (5.3 kpc) along a position angle of 65\( ^{\circ } \),
and characterized as an \ion{H}{2} galaxy by the visual spectrum of
\citet{vex}. Figure~\ref{fig:1019cont} shows the appearance of the
continuum emission for the merging system. The \( r \) band image
is of poor quality, and is included only to highlight the differing
continuum flux ratio between nuclei in the visual and infrared bands. In
the \( K_{s} \) image, the eastern galaxy appears more concentrated
than the western galaxy, which shows diffuse emission surrounding
the nucleus. While the nuclear components have a 1.6:1 flux ratio in
1\farcs 5 box apertures, the eastern nucleus being brighter, the total
continuum fluxes in larger apertures of the companion galaxies approach
the same value. Each galaxy measures \( K_{s}\approx 13.05 \) mag in \(
4''\times 5'' \) and \( 5''\times 5'' \) apertures for the eastern and
western galaxies, respectively. The \( r \) band ratio, by contrast,
is 0.7:1 in similar apertures.

\begin{figure*}[tbh]
\centerline{\epsfig{file=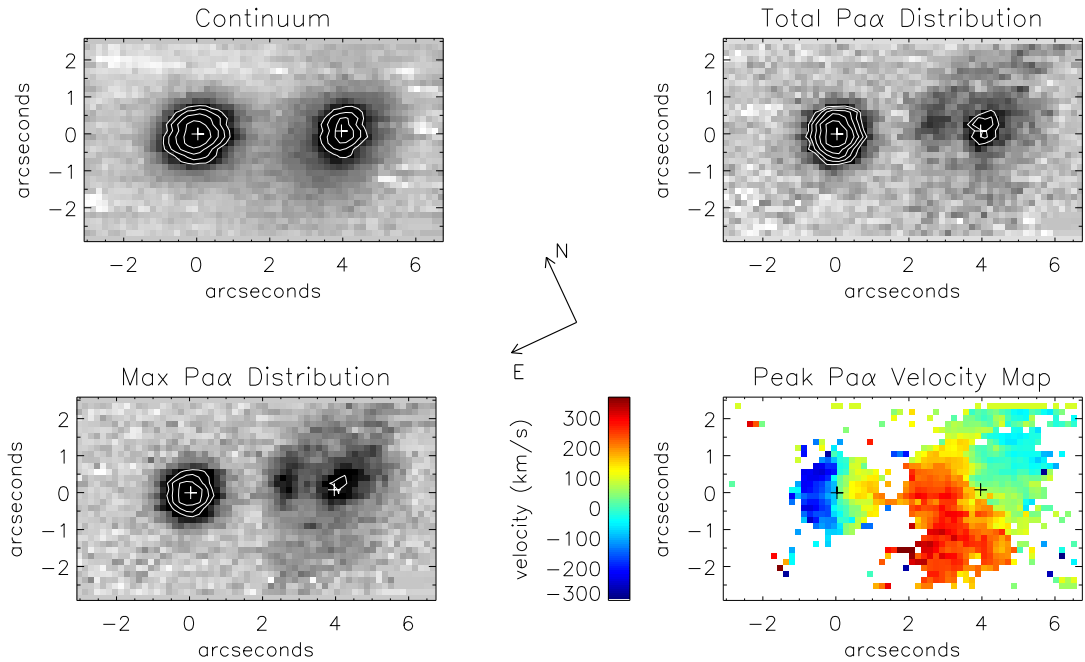,width=6.0in}}
\caption{\label{fig:1019pifs}Integral field data for
IRAS~10190$+$1322. Description of the four images follows from the
Figure~\ref{fig:0152pifs} caption. Crosses coincide with the positions of
the continuum peaks. The field orientation is as indicated by the arrows,
with ``left'' corresponding to a position angle of 65$^{\circ }$.}
\end{figure*}

Figure~\ref{fig:1019pifs} shows the appearance of the Pa\( \alpha  \)
emission across this system. The Pa\( \alpha  \) emission is distributed
similarly to the continuum emission, with a compact eastern source and a
diffuse nebula surrounding the central peak of the western source. The
ratio of Pa\( \alpha  \) fluxes in small apertures around the nuclear
concentrations is 3.2:1, while the total Pa\( \alpha  \) flux for the
two galaxies is roughly the same, indicating that the line emission is
much more diffuse in the western galaxy than in the eastern galaxy. The
off-nucleus line emission in the western galaxy is fairly uniform across
its large extent, except for a knot of emission between the two nuclei.
It is worth noting that a longslit spectrum of this pair of galaxies
would have missed the bulk of the line flux in the western galaxy,
leading one to potentially misjudge the relative importance of star
formation in these two galaxies.

\subsubsection{Star Formation \& Extinction\label{sfr1019}}

Converting the total Pa\( \alpha  \) flux in the PIFS field of \(
2.9\times 10^{-17} \) W~m\( ^{-2} \) into a star formation rate, as
described in Section~\ref{sfr0152}, yields a total system SFR of 27 \(
M_{\odot } \)~yr\( ^{-1} \), split rather evenly between eastern and
western galaxies. The clump of line emission between the two galaxies
hosts approximately 1--1.5 \( M_{\odot } \)~yr\( ^{-1} \) of star
formation. Most of the star formation is likely buried in the eastern
nucleus, as judging by the 6 cm radio continuum from \citet{crawford}. The
radio map has resolution of \( 4'' \), coinciding with the separation
of the two nuclei. The radio contours, peaking on the eastern nucleus,
are stretched out toward the direction of the western nucleus. It is
obvious that the eastern nucleus dominates the radio emission. The ratio
of radio emission to far-infrared emission in \double\ is much like that
of other starburst galaxies, with a \( q\equiv \log (S_{FIR}/S_{20\,
\mathrm{cm}}) \) parameter of 2.38, compared to the average starburst \(
q=2.40\pm 0.26 \) as reported by \citeauthor{crawford}.

Under the assumption that the radio emission traces star formation, and
that Pa\( \alpha  \) does the same, though subject to greater extinction,
one may estimate the amount of star forming activity hidden from view at 2
\( \mu  \)m in the eastern nucleus. Gaussian smoothing the Pa\( \alpha  \)
emission map to a resolution of \( 4'' \), after artificially adjusting
the flux ratio of the eastern and western contributions, one finds that
the radio map is consistent with a 2.5:1 flux ratio between the two
components. Another way to put this is that the eastern galaxy's Pa\(
\alpha  \) emission on average suffers one magnitude more extinction
than does the Pa\( \alpha  \) emission in the western galaxy. Though
no extinction estimate is available from H\( \alpha  \) measurements,
the foregoing estimate would result in a visual extinction around 7
mag. An extinction estimate based on an actual measurement of the H\(
\alpha  \) flux would almost certainly be less than this since the dust
is probably mixed with the line emission source, so that the Pa\( \alpha
\) light is incapable of probing to the center of the obscuration.

Using the extinction estimated via the 6 cm map to adjust the above star
formation rate estimate, the total SFR is closer to 50 \( M_{\odot }
\)~yr\( ^{-1} \), with much of this obscured by heavy dust extinction in
the eastern nucleus.  This is still probably a lower limit to the star
formation in the two galaxies, as extinction common to both galaxies
is not taken into account. A rough estimate of the unaccounted star
formation may be established by the expectation from \citet{scocno}
that the star formation rate will be greater than \( \sim  \)75 \(
M_{\odot } \)~yr\( ^{-1} \) for a galaxy with a bolometric luminosity
of \( 10^{12}L_{\odot } \).

\subsubsection{Merger Geometry \& Age\label{merger1019}}

\begin{figure*}[tbh]
\centerline{\epsfig{file=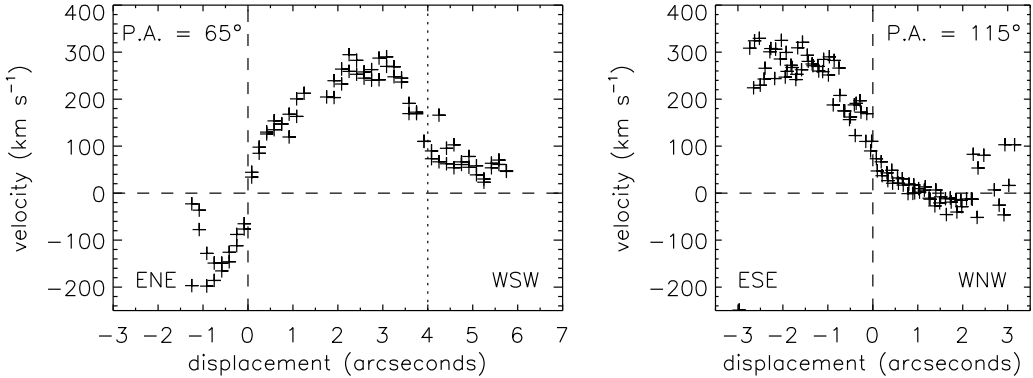,width=6.0in}}
\caption{\label{fig:1019rot}Position-velocity plots of
IRAS~10190$+$1322, with the left-hand figure representing a cut
through both nuclei simultaneously, at the field's long-axis position
angle of 65$^{\circ }$. The curve on the right is along a position
angle rotated 50$^{\circ }$ relative to the first, showing the
full velocity extent of the western galaxy. Both plots share the same
vertical scale, and the zero point is placed at the velocity of the
eastern continuum peak. The dotted line in the left-hand plot indicates
the spatial position of the western nucleus.}
\end{figure*}

Unlike the other ULIRGs in this sample, \double\ exhibits a rather
organized kinematical state---that of simple rotation of the two
components, as seen in Figure~\ref{fig:1019pifs}. The knot of Pa\(
\alpha  \) between the two nuclei does not stand out in the velocity map,
showing that it is simply a local hot-spot in the disk of the western
galaxy, and not a separate kinematical component.  Rotation curves for
each of these two galaxies are presented in Figure~\ref{fig:1019rot}.
The eastern galaxy has a total velocity amplitude of around 350 km~s\(
^{-1} \), and the more extended western galaxy shows a 340 km~s\( ^{-1}
\) velocity amplitude. The centers of these galaxies are moving at a
relative velocity of \( \sim 110 \) km~s\( ^{-1} \), with the eastern
galaxy blueshifted relative to the western galaxy. This configuration
in velocity space suggests that the eastern galaxy is in a prograde
configuration, while the western galaxy is retrograde.

The extent of the Pa\( \alpha  \) emission in the western galaxy,
as visible in Figure~\ref{fig:1019pifs}, is an impressive seven kpc
across. The morphology is suggestive of a disk inclined at about 40\(
^{\circ } \)--50\( ^{\circ } \), in which case the deprojected maximum
velocity difference along the major axis implies a circular velocity full
amplitude of 450--530 km~s\( ^{-1} \). This value is consistent with the
expectation from the \( H \) band Tully-Fisher relation \citep{tully}
based on the broadband absolute magnitude of the western galaxy (\( K_{s}=
\)13.0 mag; \( M_{K_{s}}=-24.6 \) mag; \( \Delta v\approx 530 \) km~s\(
^{-1} \), assuming \( H-K\approx 0.8 \) for ULIRGs, as in \citet{carico}).
Though it would seem that galaxies having endured a close encounter
would not be able to maintain a rotation profile in agreement with the
Tully-Fisher relation---as observed in the western galaxy---it is found
that even rather disturbed disks do not lie far from this relationship
\citep{hos97,barton}.

If indeed the eastern galaxy is prograde and the western galaxy is
retrograde, the fact that the eastern galaxy is more compact in Pa\(
\alpha  \) than is the western galaxy, and additionally is the site
of the most intense star formation, is in excellent agreement with
expectations of gas concentrations based on encounter geometry. A galaxy
in a prograde configuration experiences near resonance with the orbital
motion of the companion, leading to exaggerated tidal disturbances.
These disturbances often manifest themselves in the form of bar modes in
the galactic disks, which serve to torque and funnel gas from the disk
to the central regions of the galaxy \citep{hos96}, creating a compact
central gas concentration.  Retrograde encounters, on the other hand,
are far less disruptive \citep[see][]{toomre}, such that the disk is not
significantly disturbed owing to an averaging out of tidal forces from
the companion. Therefore, retrograde disks do not experience appreciable
nuclear gas concentrations following the initial encounter. The match to
the observational data for \double\ is reassuring, and helps strengthen
our identification of the eastern galaxy as prograde and the western
galaxy as retrograde.

Despite the fact that the majority of the infrared luminosity can
be associated with the eastern, prograde galaxy, the western galaxy
nonetheless is hosting substantial star formation activity. With
an estimated SFR of at least 13 \( M_{\odot } \)~yr\( ^{-1} \), the
western galaxy appears to have had its global star formation enhanced
by the merger activity. Though lacking a major nuclear starburst, the
western galaxy may be experiencing slight disk perturbations that lead
to cloud-cloud collisions throughout the disk, resulting in the observed
widespread distribution of star formation in this galaxy.

The organization and symmetry exhibited by the galaxies in \double\
suggests that this galaxy pair is closing in for a second encounter,
having spent a few dynamical timescales apart following the initial
encounter. This statement is based on the assumption that merging galaxies
are ultraluminous only \emph{after} the first close encounter, and that
the symmetric, orderly appearance of these galaxies can only mean that the
galaxies have had enough time to regroup after their first encounter. The
simulations by \citet{hos96}---especially the prograde/retrograde
encounters as seen just prior to the second encounter---very much resemble
the \double\ system. This scenario raises a few important issues. First,
how does one tell whether a double-galaxy merger is being seen early or
late in the merger process? Besides the organized velocity fields and
symmetric inner isophotes, what large scale morphological signatures
might there be to indicate the age of a merger? Does \double\ display
such features? The second major issue is: why now? Why wait until just
\emph{before} the final merger to become ultraluminous?

A simple statement regarding the young versus old mergers is that tidal
deformations in young mergers ought to be small and of high surface
brightness. By the time the galaxies are approaching the final merger,
tidal debris is expected to be strewn over relatively large scales, and
thus lower in surface brightness. \double\ does not, in fact, appear to
have any significant tidal structure, though deep visual imaging is not
available for this system. Inspection of the digitized Second Palomar Sky
Survey red plates does show some hint of tidal arcs around the western
galaxy, but at a level only slightly above the noise. A moderately deep
\( J \) band image taken in poor seeing conditions also shows a few
tenuous wisps of light near the western nucleus, namely at radii of 9\(
'' \) and 17\( '' \) from the western galaxy, at position angles of 205\(
^{\circ } \) and 315\( ^{\circ } \), respectively.

The question regarding why \double\ has waited until this late time in
the merger sequence to initiate ultraluminous activity is difficult to
answer. For reasons discussed at length in Section~\ref{discussion},
the ultraluminous phase must be short relative to the merger
timescale, assuming that star formation is responsible for the power
generation. Because the disks have yet to experience their second
disruptive passage, it is assumed that the observed ultraluminous
phenomenon is in response to the earlier encounter, perhaps as long
as 5--10\( \times 10^{8} \) yr ago. Encounter geometry may play a
significant role in determining the time of the onset of ultraluminous
activity, though many other factors specific to the constituent galaxies
including galaxy structure, initial gas distribution, etc. may also be
very important. If the eastern galaxy is not purely prograde, then the
weaker tidal perturbation may require more time to permit the development
of significant accumulations of gas in the nuclear regions.

\subsection{IRAS~20046\protect\( -\protect \)0623\label{ir2004}}

\subsubsection{Morphology of Continuum \& Line Emission\label{morph2004}}

\begin{figure*}[tbh]
\centerline{\epsfig{file=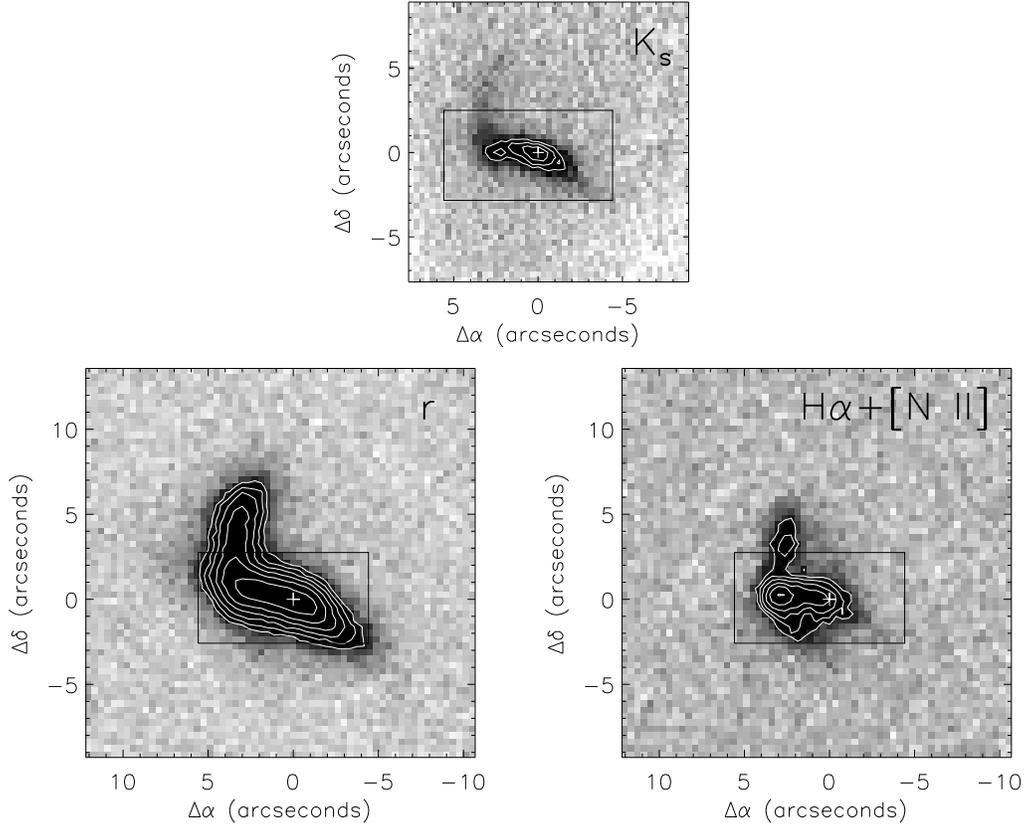,width=6.0in}}
\caption{\label{fig:2004cont}Continuum and H$\alpha
+${[}\ion{N}{2}{]} images of IRAS~20046$-$0623. The tidal tail
extending to the north is much more prominent in the $r$ band
image than in the $K_{s}$ image, suggesting that it is comprised of
young stars. The H$\alpha $ emission peaks at the eastern end of
the continuum bar, and line emission is also seen following the tidal
tail. The cross indicates the position of the near-infrared peak, and
the rectangular box represents the PIFS field. North is up, and east is
to the left.}
\end{figure*}

The morphologically peculiar \trouble , classified as a starburst
from unpublished visual spectra \citep{strausspc}, is displayed in
Figure~\ref{fig:2004cont}.  The galaxy has a continuum shape at first
appearing like a warped disk, with one prominent tidal tail extending
to the north at the eastern end of the continuum bar. The H\( \alpha  \)
image is seen to share a somewhat similar appearance to the \( r \) band
continuum, though the peak line emission is located on the eastern end
of the continuum bar. The tidal tail shows emission in H\( \alpha  \),
though displaced about 0\farcs 6 to the west of the continuum ridge.

\begin{figure*}[tbh]
\centerline{\epsfig{file=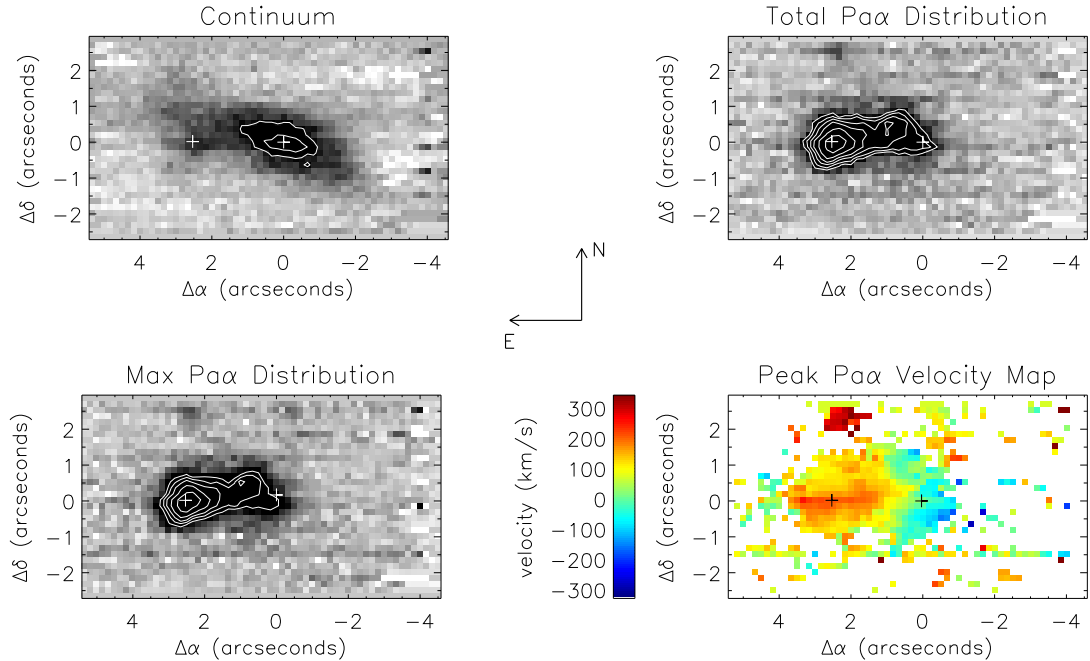,width=6.0in}}
\caption{\label{fig:2004pifs}Integral field data for
IRAS~20046$-$0623. Description of the four images follows from the
Figure~\ref{fig:0152pifs} caption. Crosses indicate the positions of
the continuum and Pa$\alpha$ peaks. North is up, and east is to
the left.}
\end{figure*}

The integral field data for \trouble\ are presented in
Figure~\ref{fig:2004pifs}.  Here, the Pa\( \alpha  \) emission and
continuum emission are largely exclusive of each other, in that the peak
of the Pa\( \alpha  \) emission coincides with the outer edge of the
continuum bar, and the peak of the continuum emission is located at the
edge of the strong Pa\( \alpha  \) structure. Faintly visible is a patch
of Pa\( \alpha  \) emission to the northeast, coinciding with the H\(
\alpha  \) emission in the tidal tail.

\subsubsection{Extinction \& Star Formation\label{sfr2004}}

The total Pa\( \alpha  \) flux in a \( 5''\times 2 \)\farcs 5 box,
oriented east-west and capturing the majority of the Pa\( \alpha  \)
emission, measures \( 3.5\times 10^{-17} \) W~m\( ^{-2} \). Comparing
this to the H\( \alpha + \){[}\ion{N}{2}{]} flux in a similar aperture
yields a comparable flux of \( 3.6\times 10^{-17} \) W~m\( ^{-2}
\). This implies a lower limit on the average extinction to the line
emitting gas of \( A_{V}>3.5 \) mag, using the same assumptions detailed
in Section~\ref{sfr0152}. A 1\farcs 5\( \times  \)1\farcs 5 aperture
centered on the peak emission yields Pa\( \alpha  \) and H\( \alpha +
\){[}\ion{N}{2}{]} fluxes of \( 1.75\times 10^{-17} \) W~m\( ^{-2} \)
and \( 7.9\times 10^{-18} \) W~m\( ^{-2} \), respectively, implying an
extinction of \( A_{V}>4.6 \) mag to this line emission.

The global star formation rate calculated from the Pa\( \alpha  \)
flux, following the method presented in Section~\ref{sfr0152}, is
about 40 \( M_{\odot } \)~yr\( ^{-1} \) for \trouble , uncorrected for
extinction. Using the average visual extinction of 3.5 mag, and taking \(
A_{Pa\alpha }=0.145A_{V} \) \citep[interpolated from the extinction law
in][]{extinction}, this value becomes \( \sim 65 \) \( M_{\odot } \)~yr\(
^{-1} \), which approaches the expected SFR for ultraluminous galaxies.

The tidal tail extending to the north also shows appreciable line
emission, as seen by the H\( \alpha + \){[}\ion{N}{2}{]} light
in Figure~\ref{fig:2004cont}.  Computing a SFR directly from this
light yields 1.4 \( M_{\odot } \)~yr\( ^{-1} \), uncorrected for
reddening. Assuming a moderate extinction of \( A_{V}=1 \) mag boosts this
to a few \( M_{\odot } \)~yr\( ^{-1} \). While not as high as the SFR in
the tail of \rubble , this is still higher than previously observed in
tidal tails \citep{hibbard96,hos98}. As with \rubble , this high rate
of star formation could be associated with the crossing orbits in very
young tails \citep[cf.][]{toomre,wallin}---consistent with the estimated
very early age of this merger, as discussed below.

\subsubsection{Merger Geometry \& Age\label{merger2004}}

\trouble\ presents a complex velocity field, as seen in
Figure~\ref{fig:2004pifs}.  There are two primary axes of importance,
which align with the major and minor axes of the continuum morphology. On
the western side of the galaxy there exists a velocity gradient with
iso-velocity contours running roughly parallel to the minor axis. This
part of the velocity field looks like rotation of the disk.  The eastern
side, where the Pa\( \alpha  \) is the strongest, has an orthogonal axis
of symmetry---along the major axis. Moreover, the velocity field in this
region is centrally peaked rather than linear in nature, with a maximum
velocity along the central axis and lower velocities to either side.

\begin{figure*}[tbh]
\centerline{\epsfig{file=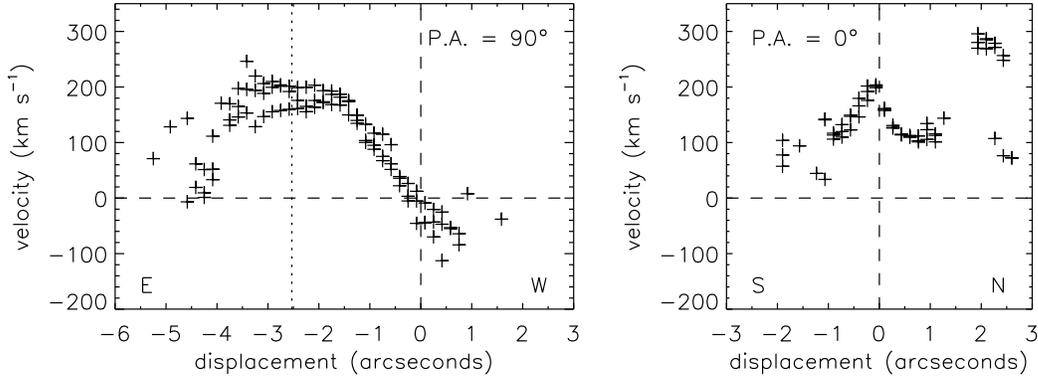,width=6.0in}}
\caption{\label{fig:2004rot}Position-velocity plots through
two different axes of IRAS~20046$-$0623.  The plot on the left
represents a 0\farcs 5 slit at a position angle of 90$^{\circ}$
through the continuum peak. The right-hand plot is along the orthogonal
direction through the Pa$\alpha$ peak. Notice the central hump at
the position of the Pa$\alpha$ peak, and the tidal tail emission
redshifted by $\sim 300$ km~s$^{-1}$ relative to the continuum
peak's central velocity. The dotted line in the left-hand plot indicates
the spatial position of the Pa$\alpha$ peak.}
\end{figure*}

The nature of the velocity fields may be seen more clearly in
Figure~\ref{fig:2004rot}, which presents position-velocity plots in
the east-west and north-south directions, centered on the continuum
peak and Pa\( \alpha  \) peak, respectively. Along the major axis,
the velocity profile has a very linear segment spanning 250 km~s\(
^{-1} \), followed by a very flat distribution across the Pa\(
\alpha  \) peak. The trailing off to the blue at the very eastern end
of the position-velocity diagram is not associated with strong Pa\(
\alpha  \) emission, and is possibly contaminated by the presence of
an OH airglow line at the wavelength corresponding to \( \sim 50 \)
km~s\( ^{-1} \). Ignoring the blue down-turn to the east, the velocity
profile of the east-west position-velocity plot as the appearance of a
flattened rotation curve, which is truncated on the western end. In the
north-south position-velocity plot, the central hump in the velocity
field is clearly seen.

The distribution of velocities in \trouble\ is not characteristic of
any simple mode of motion. However, as with \rubble , the placement and
velocity of the tidal tail feature in \trouble\ can provide some clue to
the age and geometry of this merging system. First, the relative shortness
and high surface brightness of the tail, together with the distorted
appearance of the continuum bar, argue that this system is relatively
young, being viewed soon after the first encounter.  The projected length
of the tail, roughly 7 kpc, is traveled in 3--5\( \times 10^{7} \) yr at
typical tail speeds of 150--250 km~s\( ^{-1} \). The near-straightness
of the tail, plus its high recessional velocity---seen as a cluster
of points to the north in Figure~\ref{fig:2004rot}---argues that the
parent galaxy is seen nearly edge-on, from a vantage point close to its
rotational plane. The tail, in fact, should then point at the location
of the parent galaxy's nucleus.  While there is no prominent continuum
peak along the tail centerline, the Pa\( \alpha  \) peak does in fact
lie very close to this line.

The rotational axis associated with the western continuum peak is
orthogonal to both the kinematic axis of symmetry to the east and
the rotational plane of the eastern galaxy, inferred from the tidal
tail. There are then almost certainly two distinct galaxies overlapping
to form the observed configuration. The absence of a prominent continuum
peak to the east is at first disconcerting, but the Pa\( \alpha  \) peak
belies its presence. With \( \gtrsim 5 \) mag of visual extinction to the
line emission, and therefore almost a magnitude of extinction at 2 \(
\mu  \)m, the absence of an obvious continuum peak is less startling.
The western galaxy, which by its elongation appears to be seen nearly
edge-on, could provide a screen of dust---obscuring the continuum peak
and diminishing the dominant Pa\( \alpha  \) found in the eastern galaxy.

\begin{figure*}[tbh]
\centerline{\epsfig{file=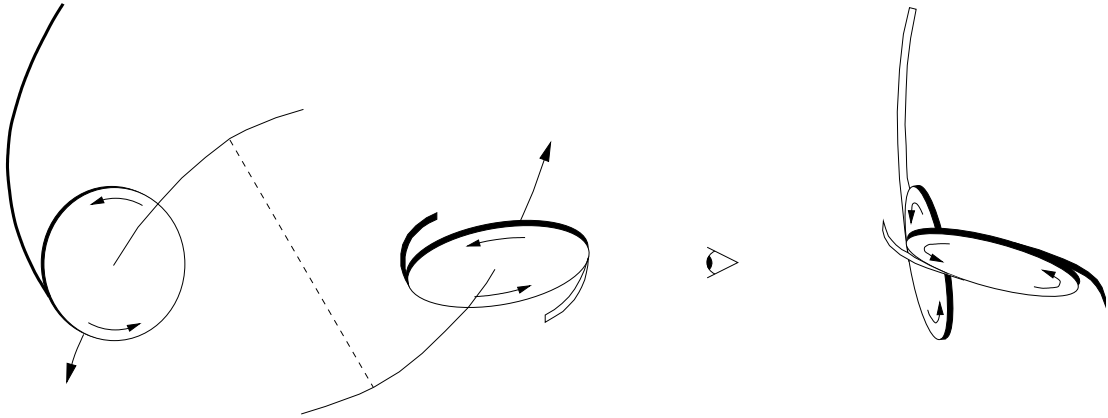,width=6.0in}}
\caption{\label{fig:ir2004geom}The proposed geometry for the
IRAS~20046$-$0623 merger, which is able to reproduce the observed
morphology and kinematics. At left is a view more-or-less onto the
orbital plane, and at right is the view we have of the system, which
is from a vantage point to the right of the left-hand figure, near the
plane of the paper. The orbital tracks and time since pericentric passage
are only suggestive here. Note that the galaxy corresponding to the one
with the bright tail and nuclear starburst is almost purely prograde in
this geometry.}
\end{figure*}

A picture can be developed that may account for the various observed
features.  Because of the extinction, we assume the western galaxy to
be in the foreground.  The western galaxy is almost edge on, so that
we are looking nearly along its rotational plane. The eastern galaxy,
mostly hidden behind the disk of the western galaxy, is also seen
more-or-less from within its rotational plane, owing to the straightness
and recessional velocity of the tidal tail. The eastern galaxy, having
a prominent tail developed rather rapidly following the first encounter,
is most likely highly prograde. If this is true, then as with \rubble ,
the eastern galaxy's rotation plane is roughly aligned with the orbital
plane, and we are looking from nearly within the orbital plane. This puts
the western galaxy in a highly inclined orbit, explaining both the lack
of a prominent tail and its tardiness at triggering any significant Pa\(
\alpha  \) emission in the nuclear region. Figure~\ref{fig:ir2004geom}
shows a depiction of the geometry of the \trouble\ system, with views
both onto the orbital plane, and from our vantage point.

The velocity field can also be understood to some degree in this
picture. As a result of its prograde geometry, and therefore its
efficiency at setting up transport of gas to the nucleus, the eastern
galaxy has highly concentrated Pa\( \alpha  \) emission associated
with its nucleus, the shape of which is seen to some extent through the
foreground screen. The rotational signature of this nucleus ought to be
red to the north and blue to the south, judging by the redshifted tail
to the north. Notice in the velocity field of Figure~\ref{fig:2004pifs}
that exactly such a gradient is seen on the southern half of the eastern
Pa\( \alpha  \) emission. The orientation of the western disk, with a
major axis position angle of \( \sim 75^{\circ } \), is placed such that
the maximum extinction would be expected just to the north of the Pa\(
\alpha  \) peak position. At this position, the diffuse emission in the
western disk may dominate the Pa\( \alpha  \) flux, masking the rotation
signature of the diminutive background line emission.  With this in mind,
the position-velocity plot on the right in Figure~\ref{fig:2004rot}
can be understood as a superposition of two components. South of the
spatial plot origin, the rotation curve of the background nucleus is seen,
but then quickly damped out to the north, where the emission from the
diffuse western disk dominates, and the associated extinction suppresses
the background flux from the eastern galaxy.

\begin{figure*}[tbh]
\centerline{\epsfig{file=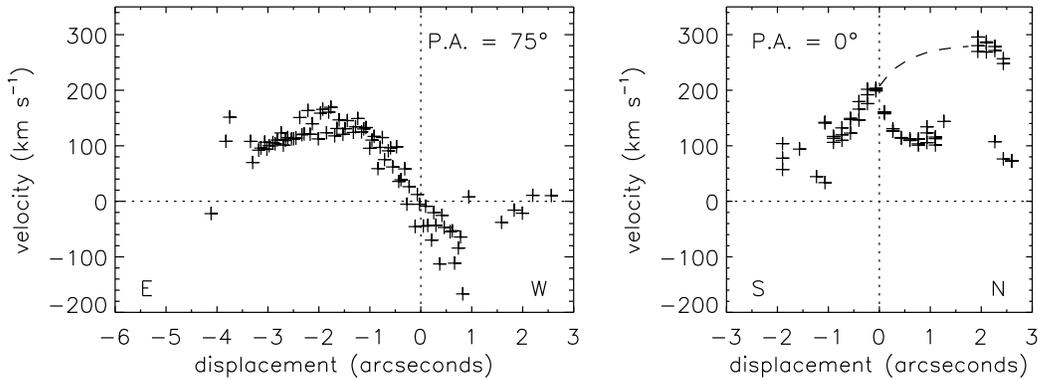,width=6.0in}}
\caption{\label{fig:2004rot2}Position-velocity plots for
IRAS~20046$-$0623, again along two axes, but this time considering
the model that has been developed for the merger geometry. The left-hand
plot now shows the rotation curve for the western galaxy, along a position
angle of 75$^{\circ }$. The right-hand plot shows the likely extension
of the rotation curve for the eastern galaxy, which is masked by the
dominant emission (plus extinction) from the western disk.}
\end{figure*}

With a better understanding of the velocity field in \trouble , a more
representative position-velocity plot can be generated for the individual
galaxies. In Figure~\ref{fig:2004rot2} we assume a 75\( ^{\circ } \)
position angle for the western galaxy, and also plot a likely extension of
the rotation curve for the eastern galaxy. Besides lying along the major
axis of the \( K_{s} \) band continuum emission, the 75\( ^{\circ } \)
position angle is in good agreement with some of the subtle features of
the velocity field in Figure~\ref{fig:2004pifs}. In particular, one sees
a demarcation between blue and yellow to the southeast of the western
nucleus that is indeed perpendicular to this position angle. Also,
the bluest emission lies to the southwest of the nucleus, rather than
directly west. The blue-colored region in the velocity field to the
north of the western nucleus does not immediately fit into the scheme,
and its nature is unknown.

Under the present geometrical interpretation, \trouble\ is a very young
interaction, being seen only a few\( \times 10^{7} \) yr after passing
through pericenter, during the first encounter. Not only does the tidal
tail morphology, velocity, and length support this idea, but the following
very simple argument strengthens this interpretation. The eastern galaxy
is highly obscured, presumably by the disk of the western galaxy---placing
the eastern galaxy in the background. The relative velocities of the
eastern and western galaxies are such that the eastern (background) galaxy
is receding, as can be seen in Figure~\ref{fig:2004rot2}.  Therefore,
the two galaxies are growing further apart, as would be expected if
these galaxies are seen just after a close encounter.

\subsection{IRAS~17574\protect\( +\protect \)0629\label{ir1757}}

\subsubsection{Morphology of Continuum \& Line Emission\label{morph1757}}

\begin{figure*}[tbh]
\centerline{\epsfig{file=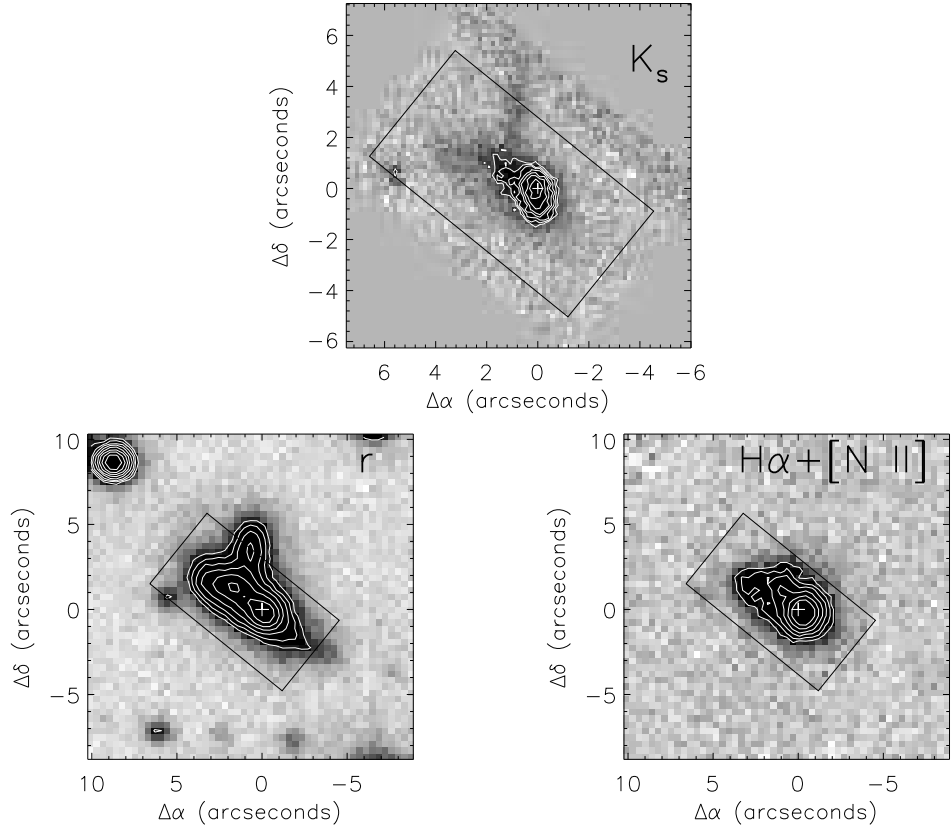,width=6.0in}}
\caption{\label{fig:1757cont}Continuum and H\( \alpha +
\){[}\ion{N}{2}{]} images of IRAS~17574$+$0629. Two tenuous tails
are seen extending to the east and north. The H$\alpha$ morphology
indicates the presence of a large, diffuse nebula to the northeast. The
rectangular box indicates the position and orientation of the PIFS
field. The cross marks the position of the near-infrared peak. North is
up, and east is to the left.}
\end{figure*}

\begin{figure*}[tbh]
\centerline{\epsfig{file=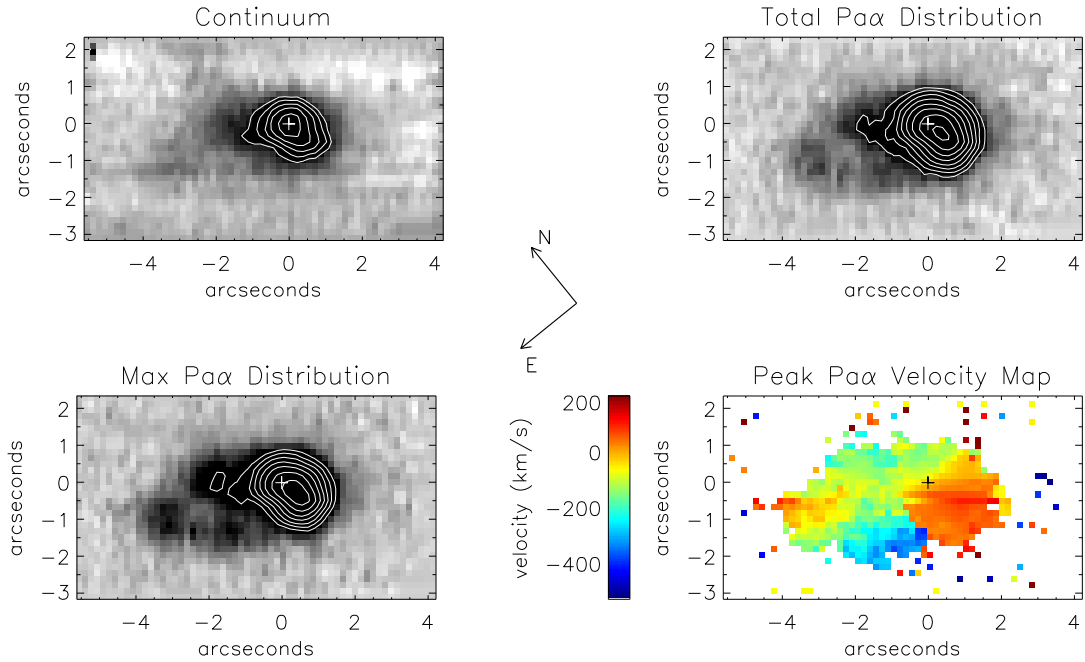,width=6.0in}}
\caption{\label{fig:1757pifs}Integral field data for
IRAS~17574$+$0629. Description of the four images follows from the
Figure~\ref{fig:0152pifs} caption. Field orientation is indicated by
the arrows, with ``left'' corresponding to a position angle of 
$51^{\circ }$. The cross indicates the position of the continuum peak.}
\end{figure*}

\bubble\ is a morphologically disturbed galaxy classified as an \ion{H}{2}
galaxy by the visual spectra of \citet{kvs}. The continuum images of
\bubble\ in Figure~\ref{fig:1757cont} show a single distorted nucleus with
what appear to be two tidal tail features extending to the north and to
the east. The H\( \alpha + \){[}\ion{N}{2}{]} image shows a prominent
emission peak slightly offset from the continuum peak, plus a large
region of diffuse emission to the northeast. Figure~\ref{fig:1757pifs}
shows the continuum as constructed from the PIFS datacube, along with
the appearance of Pa\( \alpha  \) in the system. Similar to the H\(
\alpha  \) image, a large gaseous nebula is seen in Pa\( \alpha  \)
to the northeast, roughly in the direction of the tidal features. The
detailed distribution of the Pa\( \alpha  \) nebula does not, however,
exactly trace the positions of the diverging tidal features, but rather
appears to be shaped like an edge-brightened bubble.

\subsubsection{Extinction \& Star Formation\label{sfr1757}}

The Pa\( \alpha  \) distribution can be broken into two regions: the
nuclear region, and the extended emission to the northeast. The Pa\(
\alpha  \) flux in a \( 3''\times 3'' \) box aperture around the nuclear
emission is \( 4.2\times 10^{-17} \) W~m\( ^{-2} \), while the entire
field has \( \sim  \)25\% more total flux.  Similar apertures on the H\(
\alpha + \){[}\ion{N}{2}{]} emission measure \( 6.7\times 10^{-17} \)
and \( 1.3\times 10^{-16} \) W~m\( ^{-2} \), in the same order. A lower
limit to the average extinction, using the large aperture measurement,
is \( A_{V}>2.1 \) mag, while the nuclear emission gives \( A_{V}>2.8 \)
mag. Given the flux ratios, this implies an extinction to the diffuse
emission region of \( A_{V}>0.8 \) mag. A ratio of the Pa\( \alpha  \)
and H\( \alpha + \){[}\ion{N}{2}{]} line images shows the extinction
to be separately uniform across both the diffuse region and across the
nuclear region, with typical deviations on the order of 0.2 mag.

The total Pa\( \alpha  \) emission, when converted to a global star
formation rate in the manner outlined in Section~\ref{sfr0152}, yields
103 \( M_{\odot } \)~yr\( ^{-1} \), making this the most prodigious
star forming galaxy in the sample, assuming that star formation is
indeed responsible for the Pa\( \alpha  \) emission.  This galaxy is
also slightly more infrared-luminous than the other three galaxies in
the sample. Applying the measured extinction to the Pa\( \alpha  \)
flux results in a total SFR of \( \sim 150 \) \( M_{\odot } \)~yr\(
^{-1} \), with \( \sim 85 \)\% of this coming from the \( 3''\times 3''
\) aperture around the nuclear region.

\subsubsection{Merger Geometry\label{merger1757}}

Of this sample, \bubble\ is the most difficult to understand in terms of
its status as a galactic merger. As mentioned in Section~\ref{morph1757},
the morphology of the line emission is suggestive of an intense nuclear
source accompanied by a large bubble of expanding gas. The kinematic
portrayal of the line emission in Figure~\ref{fig:1757pifs} appears quite
complex, and does not lend immediate support this picture. An expanding
bubble may be expected to display a uniform velocity gradient along the
axis of expansion, which would be to the northeast in this case. The
velocity field does not show such an orderly trend. The nuclear emission
seems to exhibit a rotation signature in the north-south direction,
but the velocity field of the diffuse emission is rather twisted and
difficult to interpret.

\begin{figure*}[tbh]
\centerline{\epsfig{file=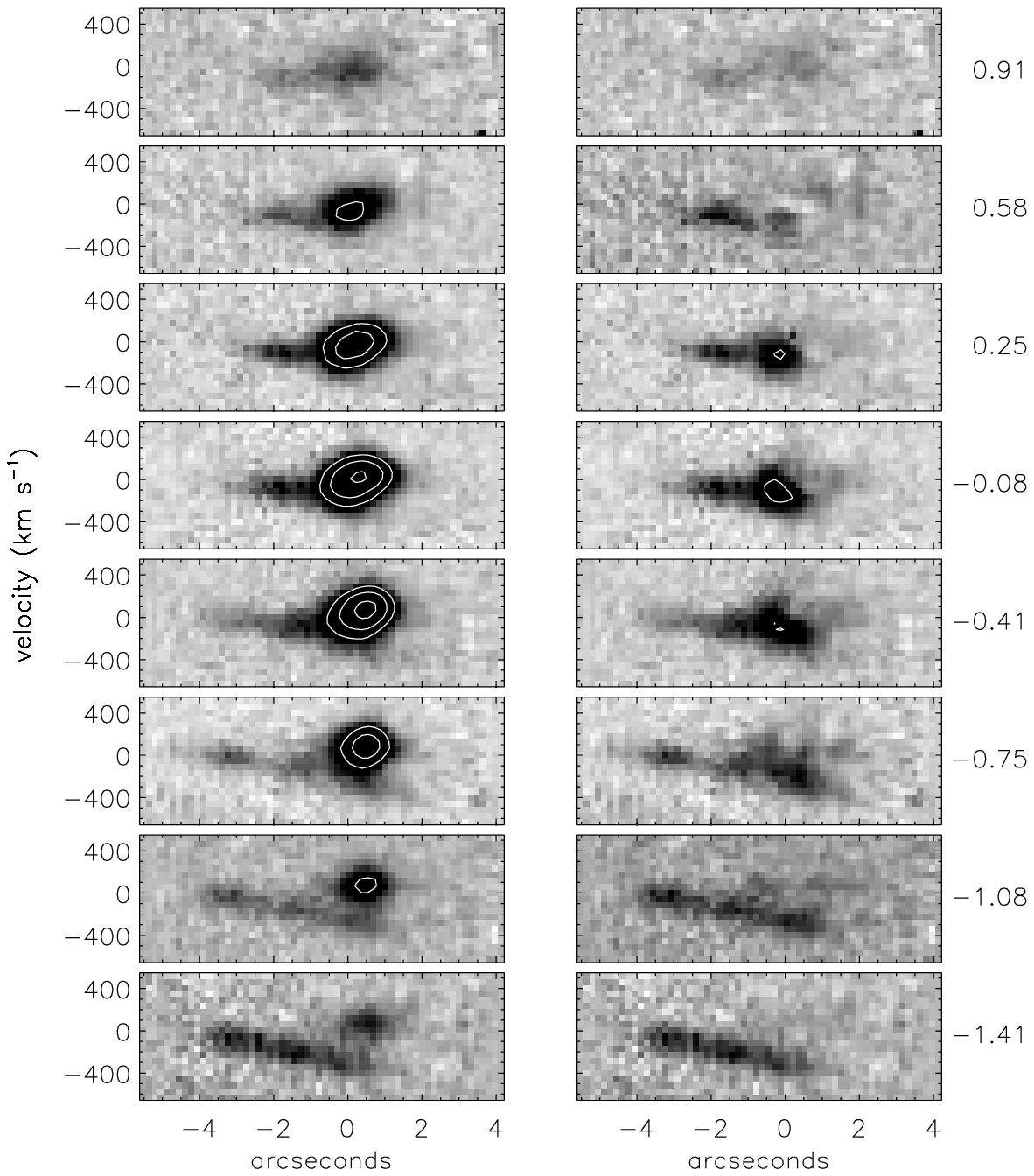,width=4.5in}}
\caption{\label{fig:17572d}A series of two-dimensional
Pa$\alpha$ spectra of IRAS~17574$+$0629 along 0\farcs 33
``slits'' at a position angle of 51$^{\circ }$.  Each panel
is spatially offset from the adjacent one by 0\farcs 33, with the
top-to-bottom progression moving from northwest to southeast. The
left-hand column depicts the continuum-subtracted total line emission,
while the right-hand column displays the same emission with the dominant
nuclear Pa$\alpha$ component subtracted. Note the multi-velocity
nature of the Pa$\alpha$ emission in the left-hand column at the
position of the nucleus. The blue component is seen to be physically
connected to the red component that appears on the left side of the
images in Figure~\ref{fig:1757pifs}. Each panel spans 10 arcseconds
horizontally and 1250 km~s$^{-1}$ vertically. Contours are placed
at multiplicative factors of two apart. The spatial coordinates
indicated at bottom and right are referenced to the coordinates used
in Figure~\ref{fig:1757pifs}.}
\end{figure*}

An interesting feature of the \bubble\ velocity field is the multi-valued
nature to the immediate southeast of the nucleus. This can be seen more
clearly in the two-dimensional spectra of Figure~\ref{fig:17572d},
especially in the three lower left panels. The red nuclear component
clearly dominates the emission, and is the component represented in
the velocity map of Figure~\ref{fig:1757pifs}, concealing the velocity
structure of the weaker blue component.

In attempting to construct a conceptual model accounting for the gas
motions, one must take the appearance of the two-dimensional spectra into
consideration, as the blue component underlying the brighter red component
seen in Figure~\ref{fig:17572d} appears to be physically associated with
the redder emission to the northeast, owing to the fact that a single,
continuous gas component smoothly connects these two regions. Note the
consistent appearance of the red-to-blue velocity gradient seen in the
faint emission of Figure~\ref{fig:17572d} in all of the synthesized slits.

\begin{figure*}[tbh]
\centerline{\epsfig{file=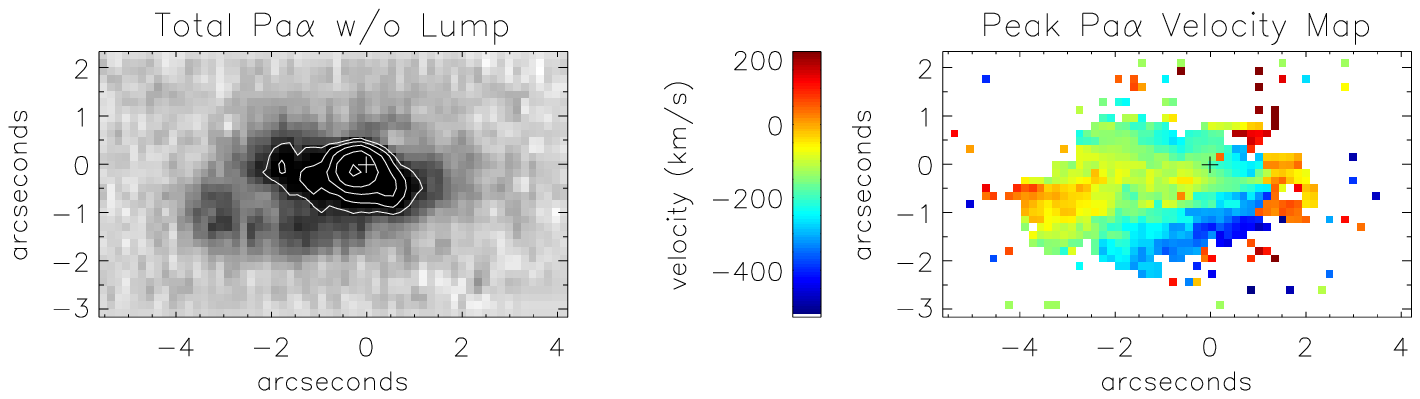,width=6.0in}}
\caption{\label{fig:1757nolump}Morphology and velocity field of
IRAS~17574$+$0629 after subtracting a mathematical representation of
the dominant nuclear Pa$\alpha$ component from the two-dimensional
spectra comprising the datacube. The underlying velocity field shows an
order that is not immediately apparent from Figure~\ref{fig:1757pifs}.}
\end{figure*}

In order to isolate this gaseous component of the system, the dominant
Pa\( \alpha  \) peak was subtracted from the datacube by fitting and
subtracting Gaussian profiles with elliptical cross sections in the
individual two-dimensional spectra. This procedure was carried out
in the two-dimensional plane represented by Figure~\ref{fig:17572d},
the result of which is also displayed here. The structure of the
Pa\( \alpha  \) nebula with the dominant peak removed appears in
Figure~\ref{fig:1757nolump}.  Judging by the appearance of the modified
two-dimensional spectra in the datacube (Figure~\ref{fig:17572d}), we
suggest that the image in Figure~\ref{fig:1757nolump} roughly corresponds
to the true morphological character of a physically connected structure
in the galaxy.

Figure~\ref{fig:1757nolump} also presents the kinematic structure of
the large Pa\( \alpha  \) nebula in the absence of the much brighter
nuclear source.  Now the velocity field appears more organized, with
a distinctive V-shaped pattern aligned with---and roughly centered
on---the nebula's major axis. Such a pattern could be consistent with
the the bubble description of the nebula, with a smooth gradient along
the expansion axis. A trend from blue to red is seen to be present
throughout the nebula as one moves away from the galaxy nucleus along
the nebula's major axis. If this feature is indeed associated with an
expanding bubble, then the observed geometry suggests that material is
being ejected somewhat towards the observer, and slowing down as it gets
farther from its source.

The bubble/outflow picture does not fully satisfy other features of
the appearance of \bubble . In particular, the \( r \) band morphology
in Figure~\ref{fig:1757cont} is fairly similar to that of \trouble\
(Figure~\ref{fig:2004cont}, Section~\ref{ir2004}).  Namely, there is a
short stub of a tidal tail extending to the north, and even the little
curl of continuum emission at the southeastern tip finds a similar analog
in \trouble ---which appears to be a very young merger experiencing its
first encounter. Indeed, the lack of low surface brightness emission
around the perimeter, much like in \rubble , argues that this too is
a first encounter ULIRG. Yet there is no obvious second galaxy. It is
possible that we are the victim of projection, wherein the second galaxy
lies just behind the visible galaxy. In fact, one clue that may support
this claim is the offset of the Pa\( \alpha  \) line emission peak from
the continuum center by about one arcsecond to the south, corresponding
to more than 1.5 kpc. It is hard to imagine the circumstances that
might produce an offset this large in a 150 \( M_{\odot } \)~yr\( ^{-1}
\) starburst. We would expect energy production on this scale to be
associated with a mass center capable of gathering the fuel supply.

Another clue in this vein may come from the Pa\( \alpha  \) image with the
dominant nuclear emission removed (Figure~\ref{fig:1757nolump}). Here, the
peak of the Pa\( \alpha  \) emission lies 0\farcs 5 to the northeast of
the continuum peak, or approximately 1 kpc away. In total, the bright
nuclear emission and the less prominent diffuse peak lie at least
2 kpc apart in the plane of the sky. Could these two emission line
features point to the locations of the parent galaxies? The continuum
shape is not entirely inconsistent with this idea. The \( K_{s} \) peak
in Figure~\ref{fig:1757cont} has a north-south elongation which could
correspond to the nuclear Pa\( \alpha  \) source---with its north-south
velocity gradient---and a spur to the northeast roughly aligned with the
strong Pa\( \alpha  \) feature as seen in the absence of the dominant
nuclear component (Figure~\ref{fig:1757nolump}).

If indeed we see two superimposed galaxies, then the tidal features
can help distinguish possible configurations. The northern tail is
almost certainly associated with the dominant Pa\( \alpha  \) peak,
as it is roughly aligned with the rotation axis of this feature. Given
the sense of rotation---redshifted to the south---the tail would be
blueshifted and in front of the galaxy. In a young encounter, this
would put the north-south galaxy in front of the more diffuse galaxy
with the northeast-southwest orientation. This arrangement goes against
the simple extinction difference between the two, though a high nuclear
extinction to the concentrated emission is by itself rather typical in
ULIRGs. However, the two-dimensional spectra in Figure~\ref{fig:17572d}
shows that the extinction to the diffuse component does not appear to be
affected by the superposition of the bright nuclear emission. This can
be seen in the three lower left panels, which shows no apparent decline
in the strength of the diffuse emission at points that are spatially
coincident with the bright component.

Determining an age for the \bubble\ encounter is impossible without a
clear understanding of its current merger state. Conflicting clues lead
to both early and late assessments. The short tails, lack of diffuse
emission surrounding the ensemble, possible double nucleus belied by
the Pa\( \alpha  \) peaks, and very close resemblance to the \trouble\
system would argue for a young merger.  The lack of two nuclei, the
incompatibility of extinction trends with attempts to reconstruct the
merger state based on kinematics, and the lack of bright line emission
in the northern tidal tail (like that seen in \trouble and \rubble )
argue that this is a late stage, post-merger system.

\section{Discussion\label{discussion}}

Ultraluminous infrared galaxies are rare, occupying the extreme tail of
the infrared galaxy luminosity function \citep{bts87,review}. Special
circumstances are required to boost the luminosity in these systems by
as much as two orders of magnitude over their normal levels. Clearly the
merging process is the key, as the overwhelming majority of ULIRGs show
evidence for recent or ongoing merger activity. An important question is:
where, and in what form is the energy produced that is being radiated at
far-infrared wavelengths as thermal dust emission?  The answer to this
question is obscured by the very dust that is radiating the far-infrared
flux.

\subsection{Physical Size \& Rate of Star Formation in ULIRGs}

In each of the ULIRGs observed in this sample, Pa\( \alpha  \) emission
is seen spread over many kiloparsecs, implying widely distributed star
formation.  But each one of these systems also exhibits a compact source
of Pa\( \alpha  \) emission with a flux comparable to---and usually
exceeding---the integrated flux of the extended line emission. The
ubiquitous appearance of a bright nuclear Pa\( \alpha  \) feature which
dominates the total star formation is an important result. Even these
spatially complex ULIRGs, selected for their extended Pa\( \alpha  \)
emission, are currently powered by concentrated nuclear starbursts. This
echoes a similar finding among a much larger sample of ULIRGs as reported
by \citet{twm01}.  Observations of ULIRGs at mid-infrared wavelengths
by \citet{bts00} and at submillimeter wavelengths by \citet{sakamoto}
also support this finding, with typical mid-infrared core sizes of \(
<200 \) pc. Also important is the fact that there typically exists at
least one magnitude of extinction at 2 \( \mu  \)m to the nuclear line
emission regions, if the total SFR is to account for the high luminosity.
Similar results have been obtained via longslit spectroscopy of ULIRG
samples \citep[][Murphy et al.]{goldader}. This is especially relevant
when searching for spectral signatures of active galactic nuclei (AGN)
within ULIRGs.

If the far-infrared luminosity in ULIRGs is completely a product of star
formation, then it is possible to calculate the rate of star formation
required to produce the observed power output. Several such estimates have
been made, establishing an order-of-magnitude expectation for ULIRG star
formation rates. The first and simplest estimate, from \citet{scocno},
computes the rate of mass consumption from O, B, and A stars via the
CNO cycle, yielding an expected 77 \( M_{\odot } \)~yr\( ^{-1} \) for
a total luminosity of \( 10^{12}L_{\odot } \). This approximation is
very nice in its simplicity, though probably an underestimate because
some of the mass consumption goes into making lower mass stars that do
not contribute significantly to the total luminosity. \citet{hunter}
integrate the stellar luminosity with an assumed initial mass function
(IMF) to obtain \( L_{ir}=10^{12}L_{\odot } \) with 260 \( M_{\odot }
\)~yr\( ^{-1} \), assuming all the luminosity is processed by dust
and emitted in the far-infrared. \citet{inoue} perform a similar, more
flexible analysis, arriving at 330 \( M_{\odot } \)~yr\( ^{-1} \), though
a rate as low as 200 \( M_{\odot } \)~yr\( ^{-1} \) can be obtained by
pushing the model to extremes. For our purposes, it is sufficient to say
that ULIRGs require an integrated star formation rate of \( \sim 200 \)
\( M_{\odot } \)~yr\( ^{-1} \).

As pointed out above, each of the ULIRGs in this sample exhibit both
widespread and concentrated nuclear star formation, as traced by Pa\(
\alpha  \) emission.  Comparison of Pa\( \alpha  \) fluxes to H\(
\alpha  \) fluxes (or 6 cm radio continuum in the case of \double )
reveals more extinction towards the sites of concentrated emission
(\( A_{V}\approx 3-7 \) mag) than to the diffuse nebular regions (\(
A_{V}\lesssim 1 \) mag). This is to be expected if the nuclear emission
arises from a very large, dense cloud of accumulated gas and dust.
The star formation rates in the extended nebular regions tend to be
10--20 \( M_{\odot } \)~yr\( ^{-1} \)---far short of the energy budget
required to produce ultraluminous emission. These SFR numbers could be
modified upward slightly by extinction corrections, though not by the
factor of ten necessary to result in ultraluminous far-infrared flux.
On the other hand, the nuclear Pa\( \alpha  \) emission translates
to typical SFR values of 50--80 \( M_{\odot } \)~yr\( ^{-1} \), after
applying extinction corrections measured from the H\( \alpha  \)/Pa\(
\alpha  \) ratio. Though this is still a factor of 3--4 short of the
ULIRG SFR requirement, it is clear that the bulk of the power comes
from the nuclear emission regions. The remaining deficit can then be
attributed to high levels of extinction in the interior of the nuclear
gas concentration, such that only the outer regions of line emission
are seen---even at the wavelength of Pa\( \alpha  \), as previously
suggested by \citet{goldader}.

\subsection{Ages of ULIRGs}

The present observations have resulted in understanding the merger
geometries of the constituent galaxies in a few ULIRGs. Knowing how the
galaxies are oriented and how they are moving relative to one another
allows one to roughly identify the time within the merger history at
which we view these systems. Measured lengths of tidal tails together
with recessional velocities yield estimates for the age of formation of
these structures. Comparisons to the gas morphologies in sophisticated
merger models lend further support to the age estimates.

The selection of the sample ULIRGs based on their spatial and spectral
complexity biases our sample in several fundamental ways. First, the
morphology biases the sample galaxies to early merger states in which
the constituent galaxies have been disrupted, yet have not coalesced
into a post-merger remnant. Also, the selection of galaxies with complex
morphologies, and whose two-dimensional spectra show large spatial extent
and non-trivial velocity structures yields a sample that presents the
greatest challenge to understand. Despite the latter effect, integral
field spectroscopy has allowed some insight into the nature of these
highly disturbed systems. One additional consequence of the selection
based partly on morphological non-triviality seems to be a greater
likelihood of chance superpositions, and alignment of our viewpoint with
the orbital plane.  This appears to play a role in at least two of the
four galaxies in the sample.

The inferred young ages of \rubble\ and \trouble\ of a few\(
\times 10^{7} \) yr, reckoned since closest approach, demand a very
fast response to the merger on the part of the gas. In order for
the ultraluminous activity to commence so quickly after the first
encounter, substantial quantities of disk gas must be very rapidly
transported into the central regions. Additional evidence for very
young ULIRGs was found by \citet{hos98} via Fabry Perot imaging of
four ULIRGs, one of which (IRAS~14348\( - \)1447) is believed to be
seen just after the first encounter. Merger models tend to show later
onsets of starburst episodes---both in terms of absolute age and with
respect to the corresponding morphological merger stage. The models by
\citet{hos96}---when scaled to Milky Way sized galaxies---show peak
star forming activity no sooner than \( 1.5\times 10^{8} \) yr after
the first encounter, corresponding to a time when the galaxies are far
apart, less distorted, and approaching their apocenter. Similarly, high
concentrations of gas do not occur in the models of \citet{barnes96}
until at least \( 10^{8} \) yr after closest approach.

In addition to the two cases of very young mergers, \double\ appears to
be seen at an intermediate age. The fact that the constituent galaxies
appear symmetric and undisturbed (i.e., lacking obvious tidal features)
supports this claim.  In order to be ultraluminous, it is assumed that
galaxies must have undergone at least one close passage---thus stimulating
the concentration of gas to fuel the starburst. When two galaxies in
an ultraluminous pair appear close together, then they have either just
passed their first encounter or are closing in for the final merger. Just
after first approach, the galaxies appear highly distorted, as seen both
in our data and in merger simulations \citep[e.g.,][]{barnes96,hos96}.
Thus the \double\ pair, with its organized morphology, has probably had
at least a dynamical timescale to regroup following the initial encounter,
and is now rapidly approaching the final merger.

As with the young mergers, \double\ exhibits a disparity between
observations and current models with regard to the time at which
merging galaxies produce energy on ultraluminous scales. The model from
\citet{hos96} ascribes an increase in the SFR relative to the pre-merger
SFR of only a factor of five at times leading up to the final merger
event when the model morphology appears strikingly similar to \double
. These models in general predict a significant burst in star formation
either following closest approach, or upon final coalescence of the
progenitor galaxies, depending on the structure of the galaxy. \double\
occupies a time in the merger sequence at which models predict a minimum
amount of star formation activity. Yet the ultraluminous status of the
galaxy pair, plus the high rates of star formation observed via Pa\(
\alpha  \) indicates by example that the ultraluminous phenomenon can
occur even at this stage in the merger sequence.

\begin{figure*}[tbh]
\centerline{\epsfig{file=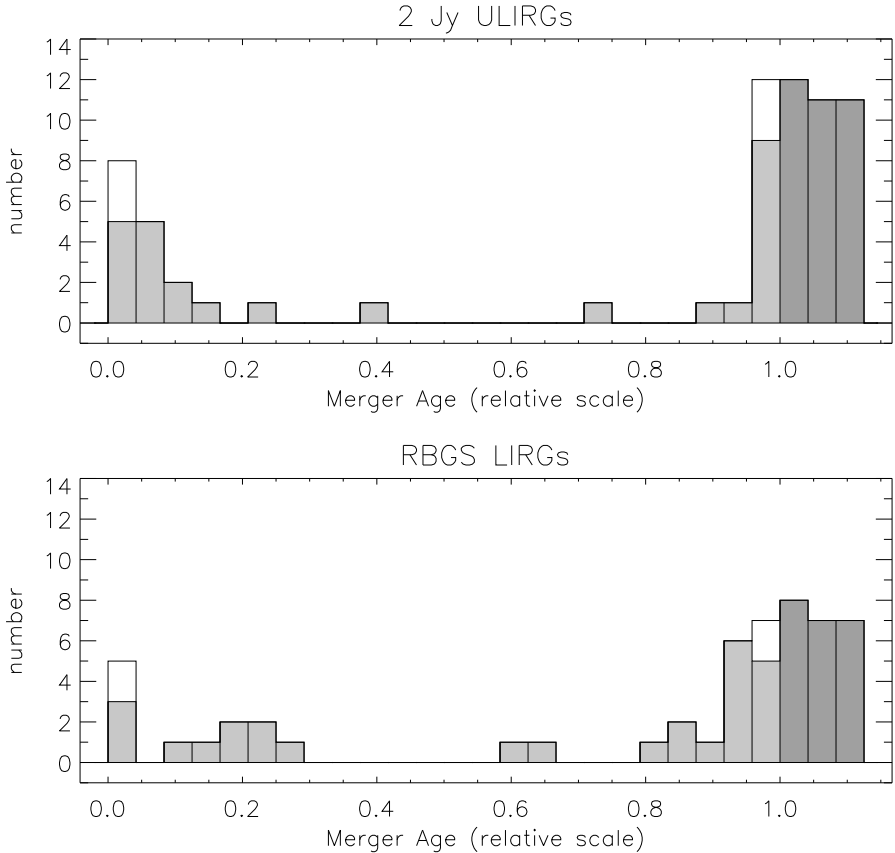,width=5.0in}}
\caption{\label{fig:hist}Histogram indicating how far ULIRGs
are along the merging sequence, as estimated from morphologies in imaging
studies. One unit of time corresponds to the encounter timescale, or
roughly $10^{9}$ yr. At the very least, ULIRGs appear to have a
bimodal time distribution, though wide separation ULIRGs indicate the
possibility that the ultraluminous phase may occur anywhere along the
sequence. The final stage ULIRGs have been arbitrarily spread across
three bins, for both scaling purposes, and also to reflect the timescale
associated with this phase. These post-merger systems are indicated
with darker colors.  The unshaded boxes represent ambiguous cases that
could not easily be classified as early or late interactions, and are
accordingly placed at both early and late times. The same information
is shown for a similar number of RBGS LIRGs, which exhibit a broader
time distribution than ULIRGs. See Appendix~\ref{append2} for details
on age classification.}
\end{figure*}

Careful study of the symmetry within ULIRG pairs in imaging surveys
suggest that ultraluminous activity just prior to a second encounter
is not entirely rare. Other ULIRGs are seen in states that appear
more consistent with a time just prior to final merger than a time
shortly after a close approach. Figure~\ref{fig:hist} shows the range of
estimated ULIRG ages, performed rather subjectively based on morphological
clues provided in imaging surveys by \citet{twm96}, \citet{sand88}, and
\citet{carico}. The time axis is cast in relative units where \( t=0 \)
corresponds to the initial encounter, and \( t=1 \) relates to the time
of final merger, roughly corresponding to \( \sim 10^{9} \) yr. For the
morphological age classification, warped isophotes and short, bright tails
signal early-stage mergers, while symmetric isophotes and large-scale
tidal debris signal late-stage mergers. Several ULIRGs were not easily
classified by this scheme, thus they appear as unshaded bins on both the
early and late sides of the distribution.  One source of ambiguity often
comes from the lack of obvious large-scale tidal structure, as this may
simply result from insufficient imaging depth. Age classifications of
double nucleus ULIRGs from the 2~Jy sample of \citeauthor{twm96} appear
in Table~\ref{tab:ages}. Details on the age classification and temporal
scaling based on nuclear separation are covered in Appendix~\ref{append2}.

These results suggest that the ultraluminous phase has a bimodal
distribution in time, with some ULIRGs found at very early times, and most
seen in the latest stages of the merger process---many of which are seen
after the progenitor galaxies have joined. A few ULIRGs are also seen at
intermediate times. The bimodality is a natural consequence of the fact
that merging galaxies on highly eccentric trajectories spend most of their
time far apart, coupled with the observation that few ULIRGs are found in
widely separated double systems. Geometry is likely to play a big role in
determining when or if ultraluminous activity occurs along the sequence
prior to the final merger. Both of the very early mergers seen in the
present sample are believed to have the ultraluminous activity associated
with highly prograde galaxies. Tilted disks, such as the northern
galaxy in \rubble\ and the western galaxy in \trouble , have some star
formation activity, mainly away from the nucleus. Perhaps these galaxies
are not sufficiently perturbed to organize a central gas concentration,
owing to the slower and weaker response of the tilted disks to tidal
disruptions. Judging by the time distribution in Figure~\ref{fig:hist},
it appears unlikely that a galaxy will enter an ultraluminous phase
during the long interval after the initial encounter, if not immediately
following first encounter. Thus highly tilted or retrograde galaxies
that require more time to concentrate their gas supplies may fail to
ever reach ultraluminous status prior to the final merger sequence.

\begin{figure*}[tbh]
\centerline{\epsfig{file=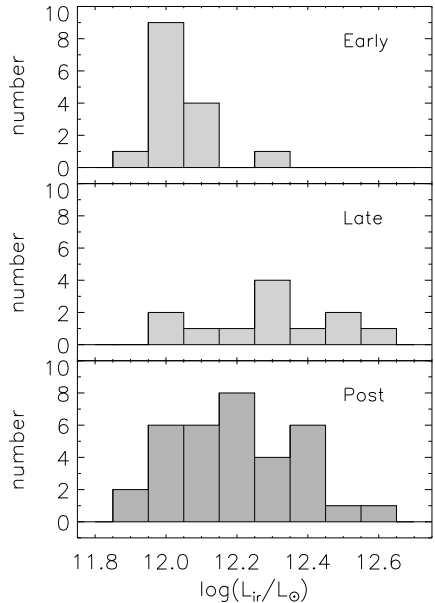,width=2.5in}}
\caption{\label{fig:lir}The distributions of infrared
luminosity for the ULIRGs identified as early-encounter, late-encounter,
or post-merger configurations.}
\end{figure*}

Though the process of assigning early or late ages to ULIRGs based
on morphology has been performed in a somewhat subjective manner,
there appears to be a significant separation in the infrared luminosity
distribution for the early population as compared with the rest of the
sample. Figure~\ref{fig:lir} shows this split in histogram form. The
late-encounter and post-merger systems are fairly uniformly spread across
the luminosity range, while early-encounter ULIRGs are seen clustered near
the low luminosity cutoff. Many previous studies of ULIRGs have associated
nuclear separation with the time remaining until final merger, and
have tried with little success to correlate properties such as infrared
luminosity, molecular gas content, etc. with this timescale. However,
grouping suspected early-encounter ULIRGs based solely upon morphology
\emph{does} tend to select lower luminosity systems than does a random
sampling, indicating that the process of of judging early or late
encounters based on morphological signatures holds some merit.  If the
age classification scheme is approximately correct, then this luminosity
separation indicates that early ultraluminous activity---usually found
occurring on only one nucleus---is accompanied by less intense star
formation activity than is typical in final merger events. This could be
because the early events do not concentrate the entire gas supply from
both galaxies into the nuclear regions in a short period of time---as is
likely the case during the final stages of a merger. The early-encounter
ultraluminous galaxies may be experiencing rather short, intense bursts
brought about by the rapid collection of \emph{some} of the galaxy's
gas content into the nuclear regions in response to the sharp tidal
perturbation caused by the recent close passage of the companion galaxy.

\subsection{Ultraluminous Timescales \& Evolutionary Scenarios}

One possibility that deserves investigation is that the ULIRG phenomenon
is episodic, with punctuated bursts of high intensity star formation,
possibly occurring more than once along the merger sequence. The idea
of repetitive starbursting was explored by \citet{noguchi}, though
in the context of cloud-cloud collisions, with most of the activity
occurring in the late stages of the merger sequence.  There exists a
fundamental problem in the assumption that merging galaxies can remain
ultraluminous throughout the entire merger process, from first encounter
to final coalescence. Defined in this way, the merger process takes 5--\(
15\times 10^{8} \) years to complete, though an ultraluminous starburst,
consuming a few hundred solar masses per year, depletes its molecular gas
supply of a few\( \times 10^{10} \) \( M_{\odot } \) in about \( 10^{8}
\) yr. Clearly then a starburst ULIRG cannot exist throughout the entire
merger lifetime. The temporal distribution in Figure~\ref{fig:hist}
strongly supports this conclusion, with a significant number of early
ULIRGs, very few intermediate-age ULIRGs, and a preponderance of
late-stage ULIRGs. The relative timescales of the starburst and merger
processes suggest an ultraluminous duty cycle around 10--20\%.

It is possible that individual mergers experience multiple ultraluminous
bursts.  Three natural time periods can be identified, corresponding
to nuclear starbursts in each of the progenitor galaxies separately,
plus a final event corresponding to the ultimate merger. Different
orbital geometries, physical structures, and gas distributions can act
to accelerate, prolong, or prevent the onset of ultraluminous activity in
each of the parent galaxies, so that independent bursts do not necessarily
happen simultaneously, if at all. As long as some appreciable fraction
of the original gas mass is preserved in the disk of at least one of
the galaxies, then it is highly likely that a final ultraluminous burst
will occur around the time of the final merger. With the possibility
of multiple ultraluminous bursts, the duty cycle for the ultraluminous
phase may be anywhere from 10--30\%.

It is entirely possible that the early bursts in ULIRGs are fueled only by
gas from the inner disk. Gas in the outer disk may not have the chance to
arrive at the nucleus before the waning tidal disruption is ameliorated by
the inherent disk stability. Alternatively, energetic outflow phenomena,
namely those produced by supernova winds, may sufficiently disrupt the
transient nuclear fueling to terminate the early ultraluminous phase.

Early ultraluminous bursts, like the ones seen in the present data, are
by no means ubiquitous among ULIRGs. Yet the fact that roughly half of
the ULIRGs are seen in separated double nucleus systems, and that roughly
one quarter are early-stage doubles according to Figure~\ref{fig:hist},
indicates that a substantial fraction of the ULIRGs pass through a
pre-coalescence ultraluminous burst. The fraction of ULIRGs experiencing
an early burst can not be ascertained directly from Figure~\ref{fig:hist}
without knowledge of the relative lifetimes of the ultraluminous phases
at early versus late times.

An important consideration to bear in mind is that ULIRGs do not have to
originate from the collision between equally gas-rich galaxies. As long
as one galaxy is gas-rich, and the two galaxies are comparable in overall
mass, the tidal perturbations can act to fuel a nuclear starburst on the
gas-rich galaxy only, or in the final merger if the first encounter is
not sufficiently disruptive.  This statement holds true if the primary
mechanism for producing ultraluminous activity stems from dense nuclear
accumulation of gas, and not from cloud-cloud collisions between the two
galaxies. Note that each of the ULIRGs observed in the present sample
appears to show true ultraluminous activity on only one galaxy nucleus,
while the companion is often only moderately active. Similar asymmetric
tendencies are observed among ULIRGs in the light of H\( \alpha  \)
\citep{armus}, and---more importantly---in quantities of molecular gas
\citep{trung}.

\subsection{Can LIRGs Be ``Resting'' ULIRGs?}

\begin{figure*}[tbh]
\centerline{\epsfig{file=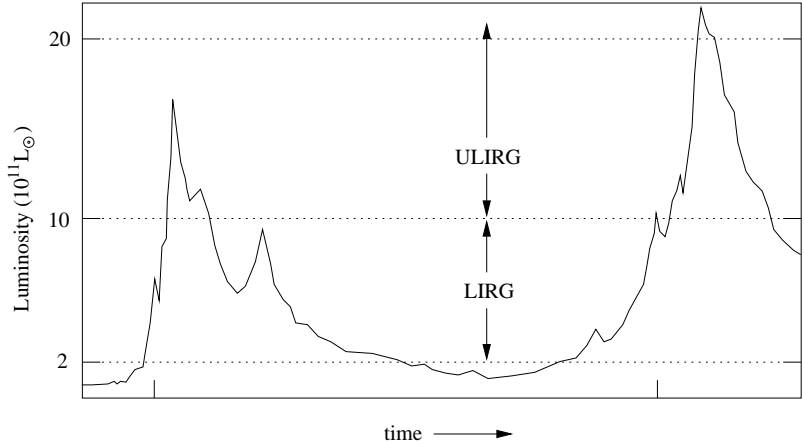,width=4.5in}}
\caption{\label{fig:cartoon}Cartoon representation of a
possible luminosity evolutionary sequence as a function of time during
the merger process. This particular example shows three principal peaks
in luminosity, corresponding to the two progenitor galaxies reacting to
the first encounter, followed later by the burst corresponding to the
final merger. In this representation, a significant portion of the merger
lifetime is spent in the ``luminous'' phase, and the ultraluminous phases
are much like islands poking above the luminosity floor. The ticks along
the time axis signify the epochs of the first and final encounters. The
fine structure in the light curve is simply intended to indicate that
critical processes associated with rates at which fuel is made available
or can collapse into stars may vary on timescales of a few million years
or less.}
\end{figure*}

LIRGs are defined as galaxies with \( 10^{11.3}L_{\odot
}<L_{ir}<10^{12.0}L_{\odot } \), with the distinction between LIRGs and
ULIRGs being an artificial boundary in luminosity. Merging galaxies may
alternate between luminous and ultraluminous states with some non-trivial
duty cycle. The problem reduces to one of luminosity evolution during the
merging process. Clearly ULIRGs pass through the ``luminous'' state during
their normal course of development. Figure~\ref{fig:cartoon} depicts an
example scenario of luminosity evolution for a major merger. This diagram
is merely suggestive, but roughly speaking, is capable of reproducing
luminosity functions for luminous infrared galaxies, as well as integrated
gas depletion via star formation.

The companion, non-active galaxies in the ULIRGs seen in this sample
provide \emph{in situ} examples of what the ULIRGs might look like in
the non-ultraluminous state. Just as observed for LIRGs, these galaxies
host extended star formation across large portions of their disks. With
moderate star formation rates of a few\( \times 10 \) \( M_{\odot }
\)~yr\( ^{-1} \), these galaxies may still preserve enough fuel supply
for a final burst, yet have enough activity to classify as LIRGs for a
substantial fraction of the encounter duration.

Indeed LIRGs are frequently found in merging pairs, with the fraction
of LIRGs classified as strongly interacting rising from 40\% at
the lower luminosity cutoff to \( \sim 100 \)\% at the high end
\citep{sandpasp}. LIRGs are more abundant than ULIRGs in a given volume
of space by at least a factor of 20--30 \citep{bts87}. If \( \sim 60
\)\% of LIRGs are associated with major mergers, then there are at least
ten times as many major mergers classified as LIRGs than as ULIRGs. If
indeed these two classes of galaxies represent different phases of the
same phenomenon, then the ultraluminous duty cycle would be around 10\%
of the luminous phase in galactic encounters between massive, gas-rich
galaxies. For the discussion to follow, only the LIRGs identified with
strong interactions are considered in comparison to ULIRGs.

Obviously this connection between LIRGs and ULIRGs is an
oversimplification.  Surely some LIRGs will never reach ultraluminous
status due to the lack of raw materials. Yet CO measurements in
interacting LIRGs and ULIRGs in similar merger states (double nucleus
systems with overlapping disks) presented by \citet{gao} indicate that
LIRGs have similar amounts of molecular gas compared to ULIRGs, with
median CO luminosities of \( 5.8\times 10^{9} \) and \( 9.0\times
10^{9} \) respectively, in units of K~km~s\( ^{-1} \)~pc\( ^{-2}
\). Yet the median infrared luminosities of the LIRGs and ULIRGs in
their sample differ by a factor of four. Moreover, carefully examining
the \citeauthor{gao} data, one finds no dramatic trend of CO luminosity
with \( L_{ir} \) across the LIRG sample.  Therefore, it seems that
interacting LIRGs and ULIRGs share similar quantities of molecular
gas. Under the conservative assumption that only half of the LIRGs
have quantities of molecular gas similar to that found in ULIRGs, the
estimated ultraluminous duty cycle changes from 10\% to 20\% based on
relative number densities. The estimates of duty cycles in this way are
consistent with the duty cycles estimated from comparing timescales of
mergers and ultraluminous starbursts.

A preliminary characterization of the morphological properties of strongly
interacting LIRGs in the Revised Bright Galaxy Sample (RBGS) \citep{mazz}
indicates that the proposed ULIRG-LIRG evolutionary connection does
hold some merit. If LIRGs were simply weaker versions of ULIRGs, then
they would be found in similar states of merging. Yet both the mean and
median separations for double nucleus LIRGs in the RBGS are roughly twice
that found in the 2 Jy ULIRG sample from \citet{twm96}.  This applies for
both samples as a whole, as well as for subsets containing only double
galaxies. The range of nuclear separations observed is the same in both
the LIRG and ULIRG samples, including the zero-separation post-merger
systems. These figures are summarized in Table~\ref{tab:seps}. The
ULIRGs in Table~\ref{tab:seps} are on average about four times more
distant than the LIRGs, so that the absence of close-separation double
LIRGs as compared to ULIRGs is a real and significant effect.

One argument against this evolutionary connection between ULIRGs and
LIRGs is that LIRGs could simply represent weaker (i.e., larger impact
parameter) galactic interactions. This would result in two observable
consequences. First, weaker interactions would obviously produce less
dramatic distortions in the way of tidal tails and warped isophotes. No
immediate difference in these morphological features is apparent in
the imaging data. Second, wider approaches would be less affected by
dynamical friction, resulting in much larger apocentric distances for
the weaker interactions. The observed range in separations among LIRGs
and ULIRGs is almost identical, so that both consequences of weaker
interactions are not directly reflected in the imaging data.

The same early/late age classification that was performed on the 2 Jy
ULIRG sample was also performed on the RBGS LIRGs, the histogram for which
is presented in Figure~\ref{fig:hist} along with the ULIRG data. The age
distributions of LIRGs are subtly different from that of ULIRGs, in that
the estimated LIRG ages do not cluster so tightly around the times of the
initial and final encounters.  Put in the context of luminosity evolution,
the age classification data are consistent with the notion that LIRGs
represent the tail of the first-encounter ULIRG burst, and also appear
as a precursor to the final merger. It should be pointed out that the
post-merger LIRGs and ULIRGs are placed into the last three bins of the
histograms in Figure~\ref{fig:hist}. In reality, there could be a radical
difference in the age distribution for post-merger LIRGs and ULIRGs.
ULIRGs are still confined temporally by the fact that they cannot produce
ultraluminous power levels via starbursts for more than \( \sim 10^{8}
\) yr. LIRGs are not quite as restricted in this regard. Studying the
morphologies of post-merger ULIRGs and LIRGs may be a way to probe the
time since merger, based on the extent to which the system has relaxed
from the violent process of coalescence. Such analysis is not attempted
here. If the LIRGs are indeed found to exist at later post-merger times
than ULIRGs, then just as with early bursts, these LIRGs could simply
be the fading embers of ULIRGs.

\section{Conclusions}

Merging galactic systems are by their very nature complex, rapidly
evolving entities. Integral field spectroscopy offers an ideal way
in which to capture this complexity across the entire two-dimensional
spatial extent. Often the added dimension of information allows one to
reconstruct the full merger geometry.  Among the more direct results of
this work are the following findings:

\begin{enumerate}
\item Ultraluminous galaxies may be found at very early times in the
merger sequence, perhaps \( \lesssim 5\times 10^{7} \) yr after the
first close encounter.

\item Even the very early events seem to produce their ultraluminous
emission from a highly concentrated nuclear starburst, in each case
associated with the prograde nucleus.

\item There exists significant extinction to the nuclear line emission
region, generally greater than 1 magnitude at 2 \( \mu  \)m. Thus a
hidden AGN would be difficult to detect, even at these wavelengths.

\item Young tidal tails can host appreciable star formation at rates
of several solar masses per year, perhaps related to the existence of
crossing orbits during the early stages of tail formation.

\item ULIRGs may also be found at intermediate times, well after the
first close encounter, and just prior to the final merger.

\item The observation that ULIRGs may be found at very young merger ages,
together with the paucity of large-separation double ULIRGs, suggests
an approximately bimodal time distribution of ultraluminous activity
within the merger sequence.  Most ULIRGs are seen just following the
first major encounter or during the process of the final merger.
\end{enumerate}

The surprising result that many ULIRGs appear to be quite young brings
into question the nature of luminosity evolution for major galactic
mergers. We suggest that many of the major galactic mergers currently
seen in less luminous states have gone through or will eventually pass
through an ultraluminous phase, depending on the efficiency and rapidity
with which molecular gas can be funneled into the nuclear regions. Most
ULIRGs seem to go through an ultraluminous state during the final merger,
but some also experience early bursts. The early-bursting ULIRGs may be
able to preserve enough fuel to sustain a second ultraluminous burst,
or alternatively, the companion galaxy may supply the raw materials for
this final event.

\acknowledgements

We are very grateful for the generosity of Joe Mazzarella, who allowed
us to access the pre-publication images of the RBGS catalog. Eiichi
Egami assisted with some of the observations. Ashish Mahabel graciously
donated telescope time to acquire a \( J \) band image of \double .
John Hibbard provided a thorough review, and many excellent suggestions,
for which we are thankful. We acknowledge Michael Strauss for his role in
the early study of the 2 Jy ULIRG sample. We also thank Gerry Neugebauer,
Christopher Mihos, Andrew Baker, Lars Hernquist, and Andreas Eckart
for helpful discussions. We thank the night assistants at Palomar, Rick
Burruss, and Karl Dunscombe for their assistance in the observations. This
research has made use of the NASA/IPAC Extragalactic Database (NED),
which is operated by the Jet Propulsion Laboratory, Caltech under contract
with NASA. T.W.M. is supported by the NASA Graduate Student Researchers
Program, and the Lewis Kingsley Foundation.  This research is supported
by a grant from the National Science Foundation.

\appendix

\section{Merger Time Classification\label{append2}}

The age of a merger, as determined from imaging data, is guided by
two principles.  The first is that tidal tails and related debris take
time to grow to large scales. When the tidal extensions are much larger
than the nuclear separation between galaxies, it may be assumed that
the galaxy pair is seen well after the initial encounter. There is, of
course, the possibility that multiple close encounters occur before the
final coalescence of nuclei. The implicit assumption in this analysis
is that only two encounters occur---an initial passage followed by a
final merger. The very presence of tidal tails indicates a short-lived
and intense disturbance in the not-so-distant past. Slow, spiraling
approaches to merging would not produce these striking features, so
that tidal tails imply moderately eccentric encounter orbits. Even
collisionless encounters drain energy and angular momentum from the
core galaxies through dynamical friction and by throwing off tidal
material to great distances. The detailed simulations of galaxy mergers
\citep[e.g.,][]{hos96,barnes96} suggest that even previously unbound
(or critically bound) galaxies with a 10 kpc first-encounter pericentric
distance undergo a single subsequent large separation before plunging back
together for the final merger. Thus any obvious and vastly extended tidal
debris is assumed to have been generated in a previous close encounter,
with the galaxies presently on the verge of permanently joining each
other. Conversely, short, high-surface-brightness tidal tails are assumed
to be recent formations, indicating a young encounter.

The second guiding morphological clue deals with the inner morphologies
of the galaxies. Because the galaxies probably spend a few dynamical
timescales between the time of first encounter and the final collision,
the galaxies have a chance to reorganize themselves, ameliorating
structural perturbations caused by the encounter. Thus highly distorted
inner morphologies, such as obvious bar modes and warped disks likely
indicate a young merger. Often, the warped features correspond to the
stubby beginnings of a tidal tail.

Large scale tidal features take precedence over the more subjective
assessment of distorted inner isophotes in assessing the age of the
merger. The absence of apparent large scale tidal structures presents an
ambiguity between their true absence or their low surface brightness. In
these cases, either the presence of a short tail-like extension or
obviously warped inner structure flags the ULIRG as an early merger. But
without these  additional clues, the age of the merger cannot be easily
determined. This population is indicated in Figure~\ref{fig:hist} by
the unshaded bins, and these bins are doubly represented at both early
and late times.

Conversion of galaxy separation to time along the merger sequence is
not performed in any sophisticated manner. The galaxies are simply
assumed to experience a quadratic separation profile in time, i.e.,
constant acceleration. It is easy to justify this behavior when the
galaxies are at apocenter, because at this time, the highly eccentric
orbits are very nearly parabolic. As the distance between the galaxies
diminishes, the acceleration would increase if the galaxies were point
masses. But the vast dark matter halos of the galaxies act to soften the
interaction. Representing the dominant mass component, the dark halos
begin to overlap, resulting in a diminished net acceleration. These
two effects roughly cancel each other, leading to a nearly parabolic
time-separation relation. Numerical integration of halo mass interactions,
using halo distribution functions like those detailed in \citet{hos96},
show little departure from this behavior.

The parameters of the parabolic separation function are chosen to
approximately represent the situation found in ULIRGs. The maximum
separation observed in the 2 Jy ULIRG sample is 48 kpc \citep{twm96},
so we will adopt 50 kpc as the nominal maximum separation for each
encounter. If the timescale for the encounter (initial close approach
to second close approach) is \( 10^{9} \) yr, the maximum relative
velocity will be 200 km~s\( ^{-1} \), closely matching observed velocity
differences in close pairs \citep{twm01}. The exact mapping function used
to convert the projected separations in Table~\ref{tab:ages} to time along
the merger sequence is \[ \Delta r=\left[ 50-2\left( \frac{t}{10^{8}\,
\mathrm{yr}}-5\right) ^{2}\right] \, \mathrm{kpc},\]
 inverting to\[
t=\left( 5\pm \sqrt{\frac{50-\left( \frac{\Delta r}{1\,
\mathrm{kpc}}\right) }{2}}\right) \times 10^{8}\, \mathrm{yr},\]
 where the plus sign is used for late stage mergers, and the minus
 sign for
early stage mergers. The resulting time stretch has the dramatic effect of
putting virtually all observed ULIRGs at either very early or very late
times. A few objects appear at intermediate times, demonstrating that
the ultraluminous phenomenon can be rather delayed. But by-and-large,
the ULIRG temporal distribution appears to be bimodal.

Table~\ref{tab:ages} presents the somewhat subjective age classifications
for the double nucleus galaxies in the complete 2 Jy sample, as defined
in \citet{twm96}.  Most of the classifications are based on the images
found in \citeauthor{twm96}, though supplemental data were obtained
from \citet{jasonwarm} and \citet{jasoncool}.  Note the tendency for
the galaxies classified as early interactions to have luminosities at
the low end of the ultraluminous range. This is displayed graphically
in Figure~\ref{fig:lir}, and suggests that this scheme of classification
is less than random.

\clearpage

\begin{deluxetable}{lcccc} 
\tabletypesize{\small} 
\tablewidth{0pt}
\tablenum{1}
\tablecaption{ULIRG Sample\label{tab:ulirgs}}
\tablehead{
\colhead{Galaxy} & \colhead{Morphological\tablenotemark{a}} & \colhead{$cz$\tablenotemark{b}} 
& \colhead{$\log\frac{L_{ir}}{L_\odot}$ \tablenotemark{c}}& 
\colhead{physical scale\tablenotemark{c}} \\
& \colhead{Classification} & \colhead{(km s$^{-1}$)} & &
\colhead{(kpc arcsec$^{-1}$)} \\
}
\startdata
IRAS 01521$+$5224 & double & 23931 & 11.95 & 1.38 \\ 
IRAS 10190$+$1322 & double & 22867 & 11.98 & 1.33 \\
IRAS 17574$+$0629 & single ? & 32701 & 12.10 & 1.82 \\
IRAS 20046$-$0623 & single ? & 25219 & 12.02 & 1.45 \\
\enddata
\tablenotetext{a}{Based on $K_s$ band morphology}
\tablenotetext{b}{Measured from this dataset at the position of the continuum peak coinciding with the strongest Pa$\alpha$ emission}
\tablenotetext{c}{Assumes $H_0=75$ km s$^{-1}$, and $q_0=0$}
\end{deluxetable}

\begin{deluxetable}{lccccc} 
\tabletypesize{\footnotesize} 
\tablewidth{0pt}
\tablenum{2}
\tablecaption{Integral Field Observations\label{tab:pifs}}
\tablehead{
\colhead{Galaxy} & \colhead{Date} & \colhead{Integration Time} & \colhead{Seeing} &
\colhead{P.A.} & \colhead{$\lambda_{{\rm Pa}\alpha}$ ($\mu$m)} \\
}
\startdata
IRAS 01521$+$5224S & 15 November 1999 & 3600 s & 0\farcs 8\phn & $87^\circ$ & 2.0248 \\
IRAS 01521$+$5224N & 15 November 1999 & 2400 s & 0\farcs 8\phn & $87^\circ$ & 2.0248 \\
IRAS 10190$+$1322 & 24\hfill March\hfill 1999 & 2400 s & 0\farcs 8\phn & $65^\circ$ & 2.0189 \\
IRAS 17574$+$0629 & 24\hfill July\hfill 1999 & 2400 s & 0\farcs 9\phn & $51^\circ$ & 2.0806 \\
IRAS 20046$-$0623 & 25\hfill July\hfill 1999 & 1200 s & 0\farcs 75 & $90^\circ$ & 2.0337 \\
\enddata
\end{deluxetable}

\begin{deluxetable}{lccc} 
\tabletypesize{\footnotesize} 
\tablewidth{0pt}
\tablenum{3}
\tablecaption{Infrared Imaging Observations\label{tab:hna}}
\tablehead{
\colhead{Galaxy} & \colhead{Date} & \colhead{$K_s$ Exposures $\times$} &  \colhead{Seeing}\\
& & \colhead{Integration Time} & \\
}
\startdata 
IRAS 01521$+$5224 & 25 November 1996 & $4\times 20$ s & 0\farcs 8\phn \\
IRAS 10190$+$1322 & 22\hfill May\hfill 1997 & $4\times 10$ s & 0\farcs 75 \\
IRAS 17574$+$0629 & \phn 3\hfill August\hfill 1996 & $6\times 30$ s & 0\farcs 7\phn \\
IRAS 20046$-$0623 & \phn 3\hfill August\hfill 1993 & $8\times 20$ s & 0\farcs9\phn \\
\enddata
\end{deluxetable}

\begin{deluxetable}{lcccc} 
\tabletypesize{\footnotesize} 
\tablewidth{0pt}
\tablenum{4}
\tablecaption{Visual Observations\label{tab:p60}}
\tablehead{
\colhead{Galaxy} & \colhead{Date} & \colhead{$r$ Exposures $\times$} &  
\colhead{H$\alpha$ Exposures $\times$} & \colhead{Seeing}\\
& & \colhead{Integration Time} & \colhead{Integration Time} & \\
}
\startdata 
IRAS 01521$+$5224 & 21\hfill January\hfill 1996 & $2\times 600$ s & $2\times 900$ s &

1\farcs 5 \\
IRAS 10190$+$1322 & \phn 9\hfill December\hfill 1996 & $1\times 300$ s & \nodata &

2\farcs 3 \\
IRAS 17574$+$0629 & \phn 3\hfill June\hfill 1994 & $1\times 600$ s & $1\times 900$ s &

1\farcs 2 \\
IRAS 20046$-$0623 & \phn 7 November 1993 & $2\times 600$ s & $3\times 900$ s &

1\farcs 5 \\
\enddata
\end{deluxetable}

\begin{deluxetable}{lcccl} 
\tabletypesize{\footnotesize } 
\tablewidth{0pt}
\tablenum{5}
\tablecaption{Merger Age Classification\label{tab:ages}}
\tablehead{
\colhead{Galaxy} & \colhead{Separation\tablenotemark{a}} & \colhead{$\log \frac{L_{ir}}{L_\odot}$} &  
\colhead{Age} & \colhead{Comments}\\
& \colhead{(kpc)}& & \colhead{Category} & \\
}
\startdata 

IRAS 00153$+$5454 & \phn 8.1 & 12.10 & early & distorted disks, confined debris \\

IRAS 01521$+$5224\tablenotemark{b} & \phn 7.7 & 11.95 & early & short, bright tails, distorted \\

IRAS 05246$+$0103 & \phn 9.6 & 12.05 & early & distorted, short tail? \\

IRAS 09061$-$1248 & \phn 6.4 & 11.97 & early & distorted disks, confined debris\\

IRAS 09111$-$1007 &36.0 & 11.98 & early & distortion, bars?---both disks \\

IRAS 09583$+$4714 &13.0 & 11.98 & early & distortion, bars---both disks \\

IRAS 10035$+$4852 &11.4 & 11.93 & early & two bright, relatively short tails \\

IRAS 10565$+$2448 &20.2 & 11.98 & early & tail length similar to separation \\

IRAS 14348$-$1447 & \phn 4.8 & 12.28 & early & short, bright tail \\

IRAS 14394$+$5332 &48.0 & 12.04 & early & near apogee---early/late irrelevant \\

IRAS 16474$+$3430 & \phn 6.5 & 12.12 & early & short, bright tail, distorted \\

IRAS 17028$+$5817 &22.5 & 12.11 & early & distorted, short tidal features \\

IRAS 18470$+$3233 & \phn 9.4 & 12.02 & early & two very distorted disks \\

IRAS 20046$-$0623\tablenotemark{b} & \phn 3.6 & 12.02 & early & short, bright tail \\

IRAS 23327$+$2913 &21.7 & 12.03 & early & distorted, young bridge/tail \\[2mm]

IRAS 00091$-$0738 & \phn 2.3 & 12.21 & late & tidal debris much larger than sep. \\

IRAS 00188$-$0856 &14.1 & 12.33 & late & symmetric, no bright tidal extensions \\

IRAS 03158$+$4227 &40.0 & 12.55 & late & both symmetric \\

IRAS 03521$+$0028 & \phn 3.6 & 12.48 & late & large, diffuse tidal features \\

IRAS 10190$+$1322\tablenotemark{b} & \phn 5.3 & 11.98 & late & symmetric, no bright tidal extensions \\

IRAS 12071$-$0444 & \phn 1.8 & 12.31 & late & tidal tail much larger than separation \\

IRAS 12112$+$0305 & \phn 3.7 & 12.27 & late & mature tidal features \\

IRAS 12540$+$5708 & \phn 2.8 & 12.50 & late & large scale tidal debris \\

IRAS 13451$+$1232 & \phn 3.9 & 12.27 & late & symmetric, large scale tidal debris \\

IRAS 15245$+$1019 & \phn 3.4 & 11.96 & late & large scale tidal tail \\

IRAS 21396$+$3623 &15.4\tablenotemark{c} & 12.40\tablenotemark{c} & late & symmetric, no bright tidal extensions \\

IRAS 22491$-$1808 & \phn 2.1 & 12.12 & late & large, diffuse tidal features \\[2mm]

IRAS 08572$+$3915 & \phn 5.8 & 12.08 & unknown & ambiguous tidal structure \\

IRAS 13539$+$2920 & \phn 7.2 & 12.04 & unknown & some distortion, could be early \\

IRAS 16487$+$5447 & \phn 5.3 & 12.12 & unknown & diffuse tidal fuzz, could be late \\

\enddata
\tablenotetext{a}{Projected separation is converted into merger age via the quadratic formulation described in Appendix~\ref{append2}.} 
\tablenotetext{b}{Member of the present sample}
\tablenotetext{c}{Revised redshift from \citet{str92} value: $cz = 44750$ km~s$^{-1}$}
\end{deluxetable}

\begin{deluxetable}{lccccc} 
\tabletypesize{\footnotesize} 
\tablewidth{0pt}
\tablenum{6}
\tablecaption{LIRG and ULIRG Separations\label{tab:seps}}
\tablehead{
\colhead{Sample} & \colhead{Number} & \colhead{Mean Separation} & \colhead{Median Separation} &  
\colhead{Separation Range} & \colhead{Mean $L_{ir}$}\\
& & \colhead{(kpc)} & \colhead{(kpc)} & \colhead{(kpc)} & \colhead{($10^xL_\odot$)}\\
}
\startdata 
LIRGs:\hfill all & 51 & 10.3 & \phn 4.9\phn & & 11.54 \\
\hfill doubles & 28 & 18.7 & 12.3\phn & 3.6--49 & 11.55 \\
ULIRGs: all & 64 & \phn 5.4 & \phn 0.15 & & 12.16 \\
\hfill doubles & 32 & 10.8 & \phn 6.4\phn & 0.3--48 & 12.13 \\
\tablecomments{The median ULIRG separation is reported as 0.15~kpc---a consequence of exactly half of the sample being measured doubles, with the minimum separation double ULIRG being Arp~220, at 0.3~kpc projected separation.}
\enddata
\end{deluxetable}

\end{document}